\documentclass[12pt,preprint]{aastex}

\let\lsim=\la
\let\gsim=\ga
\newcommand{\Deriv}[2]{\frac{\partial}{\partial #2}\,#1}

\newcommand{\zf}{z_{\rm F}}
\newcommand{\zmin}{z_{\rm min}}
\newcommand{\zmax}{z_{\rm max}}
\newcommand{\Mf}{M_{\rm f}}
\newcommand{\MB}{M_{\rm B}}
\newcommand{\MKs}{M_{\rm K_s}}
\newcommand{\Ks}{K_{\rm s}}
\newcommand{\KsAB}{K_{\rm s,AB}}
\newcommand{\KsVega}{K_{\rm s,Vega}}
\newcommand{\rK}{r_{\rm K}}
\newcommand{\Vmax}{V_{\rm max}}

\slugcomment{ApJ, submitted}

\shorttitle{Evolution of Elliptical Galaxies at $z\gsim 1$}
\shortauthors{Miyazaki et al.}

\begin{document}


\title{Evolution of Elliptical Galaxies at $z\gsim 1$ 
Revealed from 
a Large, Multicolor Sample of Extremely Red Objects \altaffilmark{1} }


\author{M. Miyazaki   \altaffilmark{2},
K. Shimasaku  \altaffilmark{2,3},
T. Kodama     \altaffilmark{4}, 
S. Okamura    \altaffilmark{2,3},
H. Furusawa   \altaffilmark{5},
M. Ouchi      \altaffilmark{2},\\
F. Nakata     \altaffilmark{4},
M. Doi        \altaffilmark{6},
M. Hamabe     \altaffilmark{7},
M. Kimura     \altaffilmark{8},
Y. Komiyama   \altaffilmark{5},\\
S. Miyazaki   \altaffilmark{5},
C. Nagashima  \altaffilmark{9},
T. Nagata     \altaffilmark{9},
T. Nagayama   \altaffilmark{9},
Y. Nakajima   \altaffilmark{9},\\
H. Nakaya     \altaffilmark{5}, 
A. J. Pickles  \altaffilmark{10}, 
S. Sato       \altaffilmark{9},
K. Sekiguchi  \altaffilmark{5},
M. Sekiguchi  \altaffilmark{8},\\
K. Sugitani   \altaffilmark{11}, 
T. Takata     \altaffilmark{5},
M. Tamura     \altaffilmark{4},
M. Yagi       \altaffilmark{4}, 
and N. Yasuda     \altaffilmark{4}
}

\email{shimasaku@astron.s.u-tokyo.ac.jp}


\altaffiltext{1}{Based in part on data collected at Subaru Telescope, 
which is operated by the National Astronomical Observatory of Japan.}
\altaffiltext{2}{Department of Astronomy, School of Science,
        University of Tokyo, Tokyo 113-0033, Japan}
\altaffiltext{3}{Research Center for the Early Universe, 
        School of Science,
        University of Tokyo, Tokyo 113-0033, Japan}
\altaffiltext{4}{National Astronomical Observatory of Japan, 
Mitaka, Tokyo 181-8588, Japan}
\altaffiltext{5}{Subaru Telescope, 
National Astronomical Observatory of Japan, 
650 N. A`ohoku Place, Hilo, HI 96720, USA}
\altaffiltext{6}{Institute of Astronomy, School of Science, 
University of Tokyo, Mitaka, Tokyo 181-0015, Japan}
\altaffiltext{7}{Department of Mathematical and Physical Sciences,
Faculty of Science, Japan Women's University, Tokyo 112-8681, Japan}
\altaffiltext{8}{Institute for Cosmic Ray Research, 
University of Tokyo, Kashiwa, Chiba 277-8582, Japan}
\altaffiltext{9}{Department of Physics, Nagoya University, 
Oizu-cho, Chikusa-ku, Nagoya 464-8602, Japan}
\altaffiltext{10}{Institute for Astronomy, University of Hawai`i, 
640 N. A`ohoku Place, Hilo, HI 96720, USA}
\altaffiltext{11}{Institute of Natural Sciences, 
Nagoya City University, Mizuho-ku, Nagoya 467-8501, Japan}


\begin{abstract}
We study the evolution of elliptical galaxies at $z\gsim 1$ 
on the basis of a sample of 247 Extremely Red Objects (EROs)
with $R-\Ks \ge 3.35$ (AB) and $\Ks \le 22.1$ (AB) 
constructed from $BVRi'z'JH\Ks$ multicolor data 
of a 114 arcmin$^2$ area in the Subaru/XMM Deep Survey Field 
taken with the Subaru Telescope and the UH 2.2m Telescope.
This is the largest multicolor sample of EROs 
covering $B$ to $\Ks$ bands ever obtained.
The threshold of $R-\Ks \ge 3.35$ is set to select passively 
evolving galaxies at $z \ge 0.8$.
By fitting template spectra of old ellipticals (OEs) and 
young, dusty starbursts (DSs) to the multicolor data, 
we classify EROs into these two classes
and estimate their redshifts.
In order to express possible star formation superposed on 
an old stellar population for OEs, the OE templates include 
combinations of a passively evolving 
bulge component and a star-forming disk component.

We find that $58 \%$ of the EROs in our sample 
belong to the OE class.
These OEs have a wide range of colors 
at any redshift, suggesting that 
OEs at $z \ge 0.8$ cannot be described by a single, 
passive evolution model with a fixed formation redshift ($\zf$).
We also find that 24 \% of the OEs are fit 
by a spectrum having a disk component 
with the $B$-band bulge-to-total luminosity ratio of $\le 0.9$, 
implying that a significant fraction of OEs at $z \ge 0.8$ 
have a non-negligible amount of star formation.

A comparison of the observed surface density of OEs with 
predictions from passive evolution models with $\zf=5, 3, 2.5$, 
and $2$ shows that the $\zf=3$ and 2.5 models reproduce 
the observed counts fairly well 
while the $\zf=5$ model and $\zf=2$ model overpredict and underpredict, 
respectively, counts at faint magnitudes.
In this paper we adopt an $\Omega_0=0.3$ and $\lambda_0=0.7$ 
cosmology.
We then derive rest-frame $B$-band luminosity functions (LFs)
of OEs in our sample at $z=1-1.5$ and $1.5-2.5$. 
We find that the LF at $z=1-1.5$ roughly agrees 
with the LF of local ellipticals if a dimming of 1.3 mag 
from $z=1.25$ to the present is assumed; 
this amount of dimming is predicted by a passive evolution with
high formation redshifts, $\zf\ge 3$. 
On the other hand, the amplitude of the LF at $z=1.5-2.5$ 
is found to be lower than that of the local LF by a factor of 
$\sim 3$ over the whole range of magnitude observed 
when a dimming predicted by the $\zf=5$ passive evolution model 
is considered. 
This deficiency becomes larger if models with lower $\zf$ are
adopted, 
meaning that the matches of the $\zf=3$ and 2.5 models 
to the observed number counts are superficial.
Taking account of a strong decrease in the number 
density of morphologically classified early-type galaxies 
at $z \gsim 1.5$ found by several authors, 
we conclude that the majority of ellipticals seen at present 
have not established either a red color or a smooth 
$1/4$-law profile before $z \sim 1.5$.

The angular correlation function of OEs shows a strong clustering.
The spatial correlation length is estimated to be 
$r_0 = 11 \pm 1 h^{-1}$ Mpc, which is larger than that for 
the present-day early-type galaxies of similar luminosities.
This suggests that OEs are selectively located in regions 
which will become present-day clusters or groups of galaxies.

Finally, we examine properties of DSs in our data.
The DSs are found to have a wide range of redshifts, 
with a peak of the distribution at $\sim 1.5$, 
and to have $E(B-V) \sim 0.4-1.2$ 
with a median of $\simeq 0.8$.
The dust-corrected $SFR$ values are 
found to span $10^{1-3}$ $M_\odot$ yr$^{-1}$.
Combined with a strong clustering seen in our DS sample, 
these high $SFR$s may suggest that DSs are progenitors of part of 
the present-day E/S0s.

\end{abstract}


\keywords{galaxies: elliptical and lenticular, cD --- 
galaxies: evolution --- 
galaxies: photometry }



\section{INTRODUCTION}
\label{sec:introduction}

The formation and evolution of early-type galaxies are 
one of the central issues of modern astronomy.
The evolution of early-type galaxies at $z \lsim 1$ has 
been relatively well constrained by many observations.
Tight color magnitude relations have been found 
in early-type galaxies, and the evolution of these relations 
has also been studied up to $z \sim 1$ 
(e.g., Bower, Lucey, \& Ellis 1992; 
Ellis et al. 1997; Kodama et al. 1998; 
Stanford, Eisenhardt, \& Dickinson 1998).
Luminosity functions of red galaxies and morphologically 
classified early-type galaxies 
up to $z\sim 1$ do not show clear number evolution 
but show mild brightening with redshift 
(Lilly et al. 1995; Brinchmann et al. 1998; Lin et al. 1999; 
Im et al. 2002).
These observations suggest that the evolution of early-type galaxies  
up to $z\sim 1$ is well approximated by the passive evolution,
i.e.,  a pure luminosity evolution after a starburst at the initial 
phase of their formation at high redshifts.
 
On the other hand, recent observations of more distant galaxies 
have revealed that early-type galaxies at $z \gsim 1$
do not appear to obey the passive evolution. 
Franceschini et al. (1998) have found a trend that 
the number density of morphologically classified 
early-type galaxies drops at $z \gsim 1.5$.
Similar results have been derived by Rodighiero et al. (2001) 
and by Kajisawa \& Yamada (2001).
Kodama, Bower, \& Bell (1999) and Menanteau et al. (2001)
have shown that a significant 
fraction of early-type galaxies in fields at $z\sim 1$ 
have colors bluer than predictions from passive evolution models.
Early-type galaxies that are blue in the rest-frame UV colors 
have also been found in clusters at $z>1$ 
(e.g., Tanaka et al. 2000; Haines et al. 2001; Nakata et al. 2001).

Extremely Red Objects (hereafter EROs), 
which are defined as objects having red optical-to-infrared 
colors such as $R-K \ge 5$ in the Vega-based magnitude system, 
are a suitable population for studies 
of elliptical galaxies at $z \gsim 1$, 
since a large fraction of EROs are thought to be 
passively evolving galaxies at high redshifts.
EROs were first discovered by the $K$-band surveys by 
Elston, Rieke, \& Rieke (1988, 1989).
Since then many authors have detected EROs
in near infrared surveys, 
although definitions of EROs are slightly different among the authors.

EROs are thought to be a mixture of mainly two different populations 
at $z \gsim 1$;  
passively evolving old elliptical galaxies (hereafter OEs) 
and dusty starburst galaxies whose UV luminosities are 
strongly absorbed by internal dust (hereafter DSs).
Both populations satisfy color thresholds for EROs, 
e.g., $R-K \ge 5$ (Vega), 
if they are located at $z \gsim 1$.
These two populations can be discriminated by their morphology 
or locations in color-color spaces.
Stiavelli \& Treu (2001) have classified 30 EROs 
on the basis of morphology of HST/NICMOS images, 
to find about 60\% of them are consistent with ellipticals.
Similarly, Moriondo, Cimatti, \& Daddi (2000) have found the fraction 
of ellipticals in their ERO sample to be $50-80\%$.
Mannucci et al. (2002) have separated the two populations 
using $R-K$ vs $J-K$ colors and have found that the two 
populations have similar abundances in their 57 EROs
(see also Pozzetti and Mannucci 2000).
To summarize, about a half or more of the whole EROs 
seem to be OEs.

EROs have been used to place constraints on the evolution 
of elliptical galaxies at $z \gsim 1$.
Barger et al. (1999) have found in a 61.8 arcmin$^2$ area 
that only a small number of galaxies have 
colors redder than $I-K=4$ (Vega), 
and inferred that only a fraction of the local elliptical galaxies 
have formed in a single burst at high redshifts.
On the other hand, McCracken et al. (2000) have detected 
about three times more EROs with $I-K \ge 4$ (Vega) 
in a field of 47.2 arcmin$^2$.
More recently, Daddi, Cimatti, \& Renzini (2000b) have found,
on the basis of a survey of a much wider area ($\sim 850$ arcmin$^2$),
a good agreement between the observed surface density of EROs 
and predictions by passive evolution models, 
emphasizing that surveys of small fields are likely to 
suffer from field-to-field variations of ERO density, 
because EROs are strongly clustered (Daddi et al. 2000a).

Since EROs are rare and clustered, wide-field surveys are essential 
for the studies of statistical properties of OEs like number density.
Furthermore, as EROs are a mixture of OE and DS populations,
it is also necessary to remove DSs from ERO samples 
in order to investigate the OE population.
Among the previous studies on evolution of elliptical galaxies 
using EROs, none is based on an OE/DS-separated sample 
from a wide ($\sim 100$ arcmin$^2$) area survey.
We have recently completed an imaging survey 
of a 114 arcmin$^2$ area 
in the $B,V,R,i',z',J,H,\Ks$ bands, and detected 247 EROs
with $R-\Ks \ge 3.35$ (AB mag).
These multicolor data enable us not only to classify the EROs into 
OEs and DSs by colors but also to estimate their photometric redshifts.
In this paper, we study the number density and colors of the OEs 
in this sample, and give strong constraints on the evolution of 
elliptical galaxies at $z \gsim 1$.

The structure of this paper is as follows.
We describe observations and data reduction in \S 2. 
The ERO sample is constructed in \S 3. 
Classification of EROs into the two populations is 
performed in \S 4, and 
results and discussion are presented in \S 5. 
The method of classification is described in Appendix.
In \S 6, we give a brief description on dusty starburst galaxies 
found in our sample.
A summary is given in \S 7.

In what follows, 
AB magnitudes are used throughout if otherwise noted. 
The Vega-based magnitudes are represented by (Vega) when necessary.
The scales between the Vega-based magnitudes and the AB magnitudes
are given by the following formulae: 
$\KsAB=\KsVega+1.8$ and
$(R-\Ks)_{\rm AB}=(R-\Ks)_{\rm Vega}-1.6$.
Here we assume the zeropoint flux of our ``Vega'' system 
$\Ks$ to be $6.9 \times 10^{-21}$ erg s$^{-1}$ cm$^{-2}$ Hz$^{-1}$.
We adopt an $\Omega_0=0.3$ and $\lambda_0=0.7$ cosmology, 
and express the Hubble constant as 
$H_0 = 100 h$ km s$^{-1}$ Mpc$^{-1}$.

\section{OBSERVATIONS, DATA REDUCTION, AND PHOTOMETRY}

\subsection{Optical Data}

We took deep $B$-,$V$-,$R$-, $i'$-, and $z'$-band imaging data 
of a central $30'\times 24'$ area in the 
Subaru/XMM-Newton Deep Survey Field 
($2^h 18^m 00^s$,$-5^\circ 12 ' 00''$[J2000]) with 
the prime focus camera (Suprime-Cam; Miyazaki et al. 2002)
mounted on the 8.2m Subaru telescope
during the commissioning observing runs on November 24-27, 2000 
and October 14, 18, 19 and November 17, 2001.
The image scale of Suprime-Cam is $0.''202$ per pixel.
The individual CCD data were reduced and
combined using IRAF and the mosaic-CCD data reduction software 
developed by us (Yagi et al. 2002).
The combined images for individual bands were aligned and 
smoothed with Gaussian kernels to match their seeing sizes. 
The $B$, $V$, $i'$ data and part of the $R$ data (58min), 
all of which were obtained in 2000, have already been reduced 
by Ouchi et al. (2001) for studies 
of Lyman Break Galaxies at $z\sim 4$.
Hence, in this study we reduced the $z'$ and 
newly added $R$ data alone.
The final images cover a contiguous
618 arcmin$^2$ area with a PSF FWHM of $0.''98$.
Photometric calibrations are made using photometric standard 
stars given in Landolt (1992) for $B,V,R$ data 
and spectrophotometric standard stars given 
in Oke (1990) for $i'$ (SA95-42) 
and in Bohlin, Colina, \& Finley (1995) 
for $z'$ (GD71).
Transformation from Vega-based magnitudes to AB magnitudes 
is made following Fukugita, Shimasaku, \& Ichikawa (1995).
The photometric zero points for $B$, $V$, and $i'$ data are the same 
as those adopted in Ouchi et al. (2001).
A summary of the optical data is given in Table \ref{tab:Obsdata}.

\subsection{Infrared Data}

Wide-field $J,H,\Ks$ data were obtained with 
the University of Hawaii (UH) 2.2m Telescope at Mauna Kea 
on the nights of August 28 to September 4, 2001.
The observations were made using 
Simultaneous 3-color InfraRed Imager for Unbiased Survey 
(SIRIUS; Nagayama, Nagashima, Nakajima et al. 2002).
SIRIUS is a three-color simultaneous camera
employing three $1024 \times 1024$ HgCdTe arrays. 
The field of view at each band is $4.'7 \times 4.'7$, 
with a pixel size of $0.''28$.
SIRIUS was developed as an infrared imager for the 
Nagoya University 1.4m telescope 
at South African Astronomical Observatory (SAAO), 
but it can also be mounted on the UH 2.2m Telescope.

\begin{figure*}
\epsscale{1.0}
\plotone{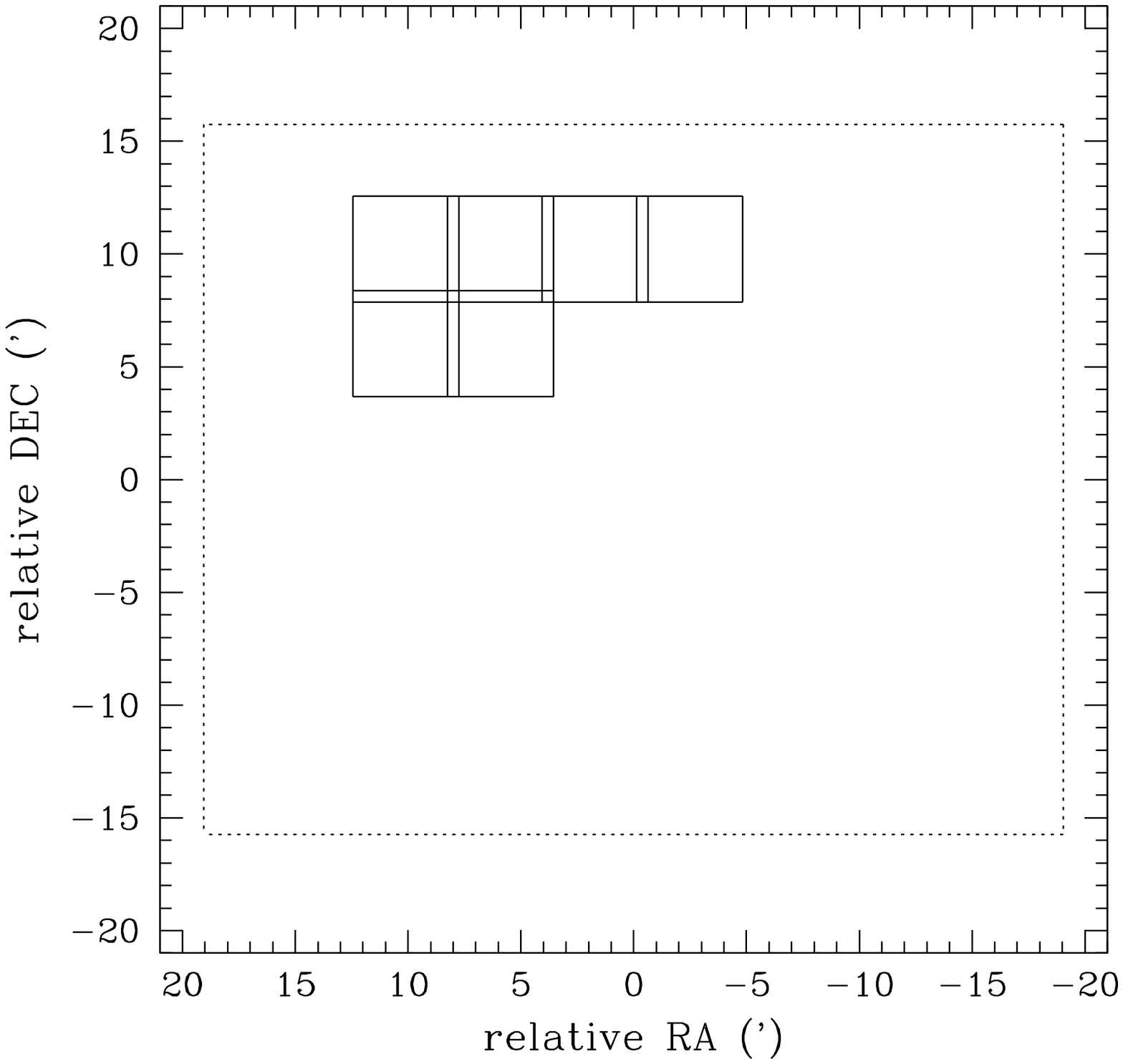}
\caption{Part of the Subaru/XMM-Newton Deep Survey Field.
The field observed by Suprime-Cam ($B,V,R,i',z'$) is 
         outlined by dotted lines, 
         and the fields (six pointings) 
         for which $J,H,\Ks$ data were taken
         by SIRIUS are outlined by solid lines.
    \label{fig:field}}
\end{figure*}

Six pointings (133 arcmin$^2$ in total) were made with SIRIUS 
in the $618$ arcmin$^2$ region imaged by Suprime-Cam.
For each pointing, we obtained 120 dithered exposures, 
each being 1 minute long.
The net exposure times of $J, H$, and $\Ks$ images 
for a single pointing are thus 120 min each.
$J$ data were not successfully acquired for a quarter of the 
field of view due to a detector problem. The area covered by $J$ 
is thus 75\% of that covered by $H$ and $K$. Figure \ref{fig:field} 
shows the regions on the sky of our optical and infrared observations.

The reduction of the $J,H$, and $\Ks$ data are carried out using IRAF.
Dome flat-fielding and sky subtraction with a median sky
frame are applied.
The reduced data are then aligned to the Suprime-Cam optical images.
The typical positional accuracy of the $J,H$,and $\Ks$ images 
relative to the optical ones is estimated to be 0.2 arcsec 
using common stars.
The final image covers a 114 arcmin$^2$ area 
for $H$ and $\Ks$ and a 77 arcmin$^2$ area for $J$, 
both having a PSF FWHM of $0.''98$, 
the same value as for the optical images.
Photometric calibrations are made 
using standard stars of Persson et al. (1998), 
and we do not make any color transformation between their 
system and the SIRIUS system.
A summary of the infrared data 
is also given in Table \ref{tab:Obsdata}.

We examine photometric accuracies of our optical $+$ NIR 
data using stars.
We compare colors of stars in our images with those of 175 bright stars 
in the spectrophotometric atlas by Gunn \& Stryker (1983)
calculated using our system response functions.
We find that for any combination of two colors 
a systematic deviation of the locus of our stars 
from that of Gunn \& Stryker's in the two-color plane 
is less than $\simeq 0.05$ mag.

\subsection{Object Detection and Photometry}

Object detection and photometry are 
made using SExtractor version 2.1.6 (Bertin \& Arnouts 1996).
The $\Ks$-band image is chosen to detect objects. 
If more than 5 pixels whose counts are above 2 $\sigma_{sky}$ 
are connected, they are regarded as an object.
In total 1308 objects are detected down to $\Ks=$22.1 mag.

For each object detected in the $\Ks$ image, 
Kron magnitudes are measured for all the bands. Our Kron magnitude,
$m_K$, is defined as
$ m_k = -2.5 {\rm log} L(<2.5\rK) $, 
where $L(<2.5\rK)$ is the luminosity within the circle of 
radius $2.5\rK$ and $\rK$ is the Kron radius (Kron 1980) 
in the $\Ks$ band.
The $\rK$ value for the Kron magnitude is 
different for different galaxies.
The fraction of $L(<2.5\rK)$ 
is about $\sim 96$\% of the the total luminosity
for both galaxy and star profiles convolved with Gaussian seeing 
(Bertin \& Arnouts 1996). 
We apply the Kron radius measured in the $\Ks$ image
to images in other bands to ensure that the colors are
measured within the same aperture for any object.

All magnitudes are corrected for Galactic absorption 
using Schlegel, Finkbeiner, \& Davis (1998), 
though the amount of absorption is quite small: 0.08 mag for $B$.

\subsection{Star-Galaxy Separation}

\begin{figure*}
\epsscale{1.0}
\plotone{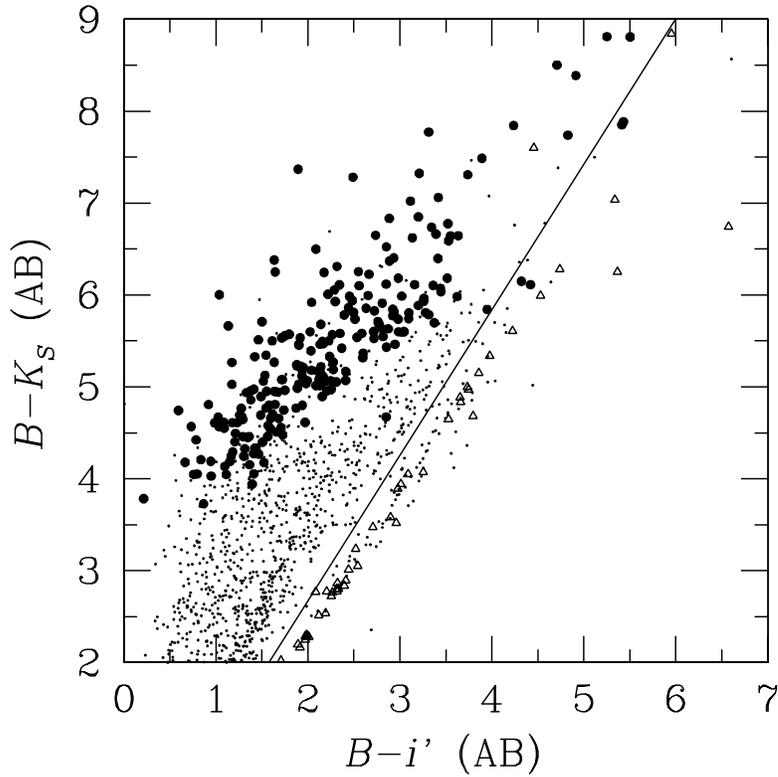}
\caption{Star/galaxy separation using $B-\Ks$ and $B-i'$ colors.
         Objects in the $\Ks$-limited sample are shown by dots 
         and large filled circles.
         Large filled circles indicate EROs, 
         and triangles correspond to 175 stars given in 
         Gunn \& Stryker (1983).
         The solid line, $B - \Ks = 1.583(B-i') - 0.5$,
         denotes the boundary between stars and 
         galaxies adopted in this study; 
         an object is regarded as a star, if it is located 
         in the right-hand side of this line and its FWHM is
         $\le 1.''2$.
    \label{fig:sgsepa}}
\end{figure*}

Star-galaxy separation is made on the basis of 
$B-i'$ and $B-\Ks$ colors and FWHM of objects.
Figure \ref{fig:sgsepa} 
plots $B-\Ks$ against $B-i'$ for all the objects 
in the $\Ks$-limited sample (dots), together with 175 stars in Gunn \& 
Stryker's (1983) spectrophotometric atlas (open triangles).
On the basis of this figure, we regard an object as a star 
and remove it from the sample, if it satisfies simultaneously
$B - \Ks < 1.583(B-i') - 0.5$ and FWHM$ \le 1.''2$.
The threshold of FWHM is imposed so that clearly non-stellar
extended objects are not removed irrespective of their colors.
In total 95 objects are regarded as stars and removed.

Large filled circles in Figure \ref{fig:sgsepa}
denote EROs to be selected 
by $R-\Ks$ color in the next section.
The distribution of EROs is well separated from the stellar 
sequence, and thus it is expected that the ERO sample 
does not suffer from either the incompleteness 
due to misidentifying EROs as stars, 
or the contamination due to misidentifying stars as EROs.

\section{THE ERO SAMPLE}

We define in this paper EROs as objects whose $R-\Ks$ color 
is equal to or redder than $3.35$, 
or equivalently 4.95 in the Vega-based magnitude. 
The threshold value corresponds to the apparent color 
of a passively evolving galaxy at $z=0.8$ 
predicted on the basis of Kodama \& Arimoto's (1997; KA97) 
population synthesis models.
In more detail, this model galaxy, being formed at $z=5$, 
has the slope of the initial mass function (IMF) of $x=1.10$,
the $e$-folding star formation timescale of $\tau_{\rm SF}=0.1$ Gyr, 
the $e$-folding gas infall timescale of $\tau_{\rm infall}=0.1$ Gyr, 
and the age when the galactic wind blows (ie, no star formation 
occurs after this) of $t_{\rm GW}=0.2$ Gyr.
This model reproduces well the average colors of massive elliptical 
galaxies observed at $z\lsim1$ (Kodama et al. 1998).

\begin{figure*}
\epsscale{1.0}
\plotone{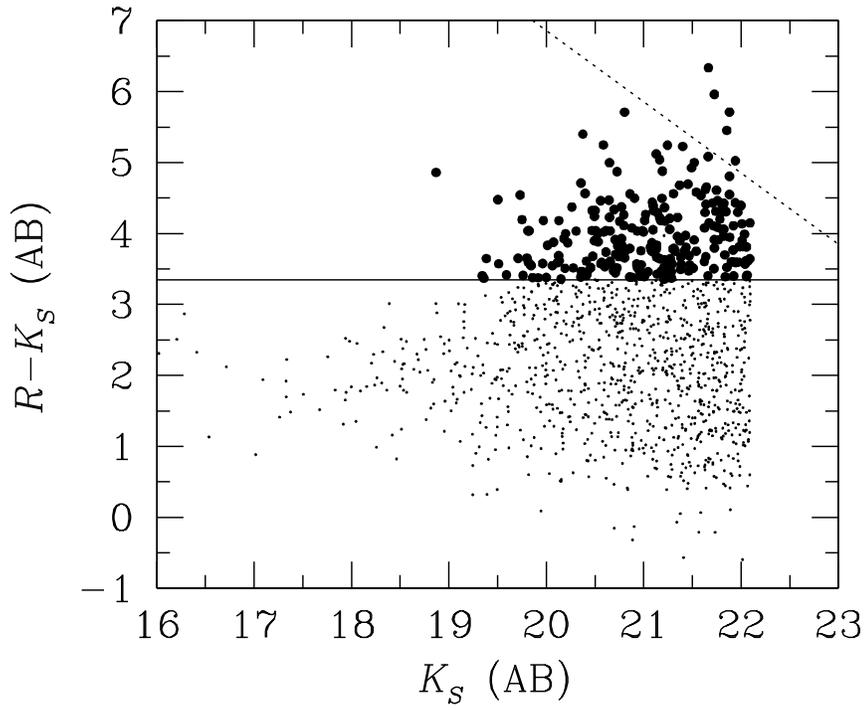}
\caption{$R-\Ks$ plotted against $\Ks$ for all objects 
         except for stars in the $\Ks$-limited sample.
         Large circles denote EROs, and the horizontal line 
         corresponds to the threshold ($R-\Ks =3.35$) 
         for selecting EROs. The dotted line indicates 
         the $3\sigma$ limiting magintude of $R$ ($26.9$).
    \label{fig:RK_K}}
\end{figure*}

Figure \ref{fig:RK_K}
plots $R-\Ks$ as a function of $\Ks$ magnitude 
for all objects except for stars in our $\Ks$ sample. 
The horizontal line denotes our boundary for the ERO selection, 
and the dotted line corresponds to the $3 \sigma$ detection 
limit for $R$ magnitudes.
There are 247 objects which satisfy the ERO threshold, 
$R-\Ks \ge 3.35$, down to $\Ks=22.1$ mag. 
We refer to these objects as the ERO sample.

We estimate the completeness and contamination of our ERO sample
as a function of apparent magnitude by Monte Carlo simulations. 
We generate 2500 objects that mimic 
the distribution in the $R-\Ks$ vs $\Ks$ plane 
of objects with $\Ks\le 22.1$, 
and distribute them randomly on our original $R$ and $\Ks$ images
after adding Poisson noises according to their original brightness.
We here assume the shapes of generated objects to be Gaussian, 
with their FWHM values scattered over the range of real 
objects, $\sim 1''-2''$.
Then, we run the SExtractor in the same manner 
as for the original images, detect these simulated objects, 
and measure their brightness, if detected.
We define here the completeness of the ERO sample as the number of
the simulated objects which again pass the color threshold 
divided by all the original objects which passed the threshold.
The contamination of the sample is defined as the number ratio 
of the objects that did not pass the threshold in the original
data but satisfy the threshold in the simulated data,
due to photometric noises, 
divided by the number of all the original objects which passed
the threshold.
The differential completeness is found to drop to $50\%$ 
at $\Ks \simeq 21.6$ mag. 
The 'cumulative' completeness down to $\Ks = 22.1$, 
which is the completeness 
weighted by the number of the detected objects, is 
estimated to be $60\%$.
The contamination is found to be lower than $10\%$ over the whole 
magnitude range.

\begin{figure*}
\epsscale{1.0}
\plotone{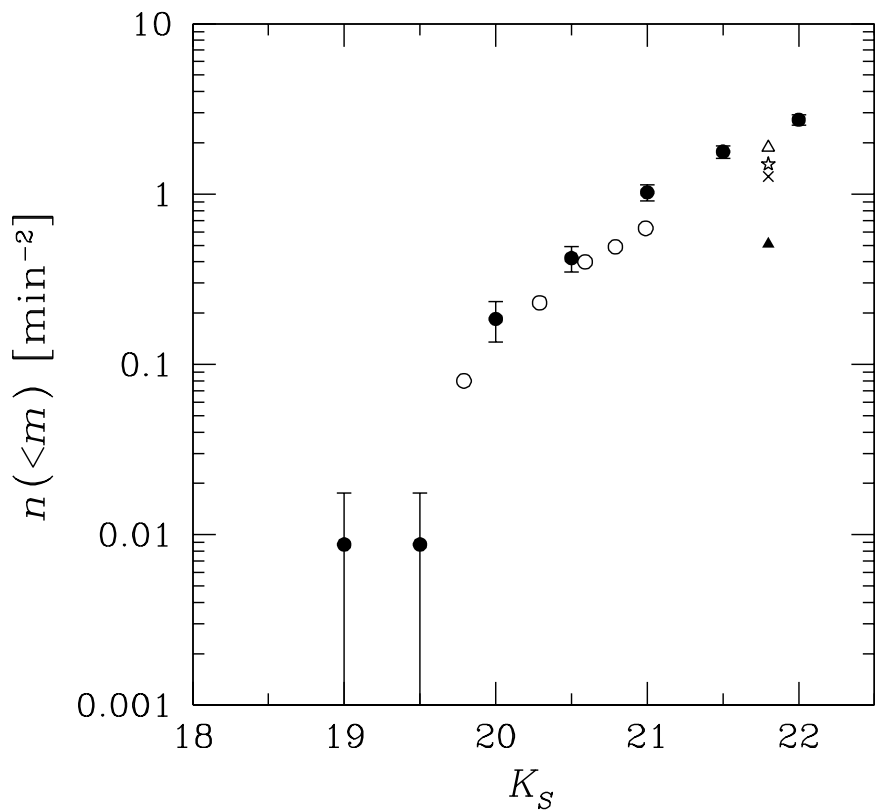}
\caption{Cumulative number counts of EROs.
         The filled circles denote our results.
         The data taken from the literature are also shown: 
         Daddi et al. (2000a; based on an area of 701 arcmin$^2$; 
         open circles), 
         Thompson et al. (1999; 154 arcmin$^2$; open triangle), 
         Cimatti et al. (2002; 52 arcmin$^2$; star), 
         Scodeggio \& Silva (2000; 43 arcmin$^2$; filled triangle),
         and Cohen et al. (1999; 14.6 arcmin$^2$; cross), 
         respectively. 
    \label{fig:nmERO}}
\end{figure*}

Figure \ref{fig:nmERO}
shows cumulative number counts of EROs in our sample 
after correction for completeness and contamination, 
together with those taken from the literature.
The filled circles denote our results.
The open circles, open triangle, star, 
filled triangle, and cross are
the counts taken from 
Daddi et al. (2000a; based on an area of 701 arcmin$^2$), 
Thompson et al. (1999; 154 arcmin$^2$), 
Cimatti et al. (2002; 52 arcmin$^2$), 
Scodeggio \& Silva (2000; 43 arcmin$^2$),
and Cohen et al. (1999; 14.6 arcmin$^2$), 
respectively.
We find from this figure that our counts match reasonably 
well with Daddi et al.'s (2000a) over $20 \lsim \Ks \lsim 21$ 
and with Thompson et al.'s (1999) 
and Cimatti et al.'s (2002) at $\Ks = 21.8$.
Cohen et al. have given a bit lower count at $\Ks=21.8$, 
but the difference is within Poisson errors.
Scodeggio \& Silva (2000) have obtained a significantly smaller count, 
which might be due to field-to-field variations.

\section{CLASSIFICATION OF EROs USING THEIR MULTICOLORS}

In this section, we classify our EROs into two classes, 
passively evolving old ellipticals (OEs) and dusty starbursts (DSs), 
using $BVRi'z'JH\Ks$ colors.

\subsection{Brief Summary of the Method}

The method of our classification is a variant of the photometric 
redshift technique. 
We use model spectra computed on the basis of stellar population 
synthesis models by KA97. 
A major difference from the ordinary 
photometric redshift technique (e.g., Furusawa et al. 2000) 
is that the model spectra we use are restricted 
to those appropriate to OEs and DSs.
The model spectra of OEs include those with some ongoing star formation
as well as those of passively evolving population.

First, a set of model spectra for OEs and DSs 
over the redshift range of $0 \le z \le 4$ are generated.
Effect of dust extinction is taken into account.
Next, the spectra are convolved with the system response functions
to give the magnitudes to be observed in the bands used in this study.
In this process, the effect of absorption due to intergalactic
HI gas is taken into account.
Then, observed magnitudes of an ERO are compared with model magnitudes 
and the $\chi^2$ value is computed for all the models.
Finally, for each ERO, we identify the model which gives the 
smallest $\chi^2$ value and assign the class of that spectrum 
to the ERO.

Out of the 247 EROs in our sample, 143 are classified 
as OEs while 104 are classified as DSs. 
Our method gives the estimated photometric redshift and some 
parameters useful to investigate systematic properties of EROs
in addition to their classifications.
Details of the model spectra and the technique are given 
in Appendix.

\subsection{Comparison with Classification Based on 
$J-\Ks$ vs $R-\Ks$ Colors}

Pozzetti \& Mannucci (2000) have introduced a method to 
classify EROs into OEs and DSs on the basis of 
their locations in the $J-K$ vs $R-K$ plane.
This simple method makes use of a characteristic difference 
in the spectra of OEs and DSs located at  $1 \le z \le 2$; 
OEs have a sharp spectral break around 4000\AA{\hspace{2pt}} while 
DSs' spectra are smoother, giving DSs' $J-K$ colors redder than OEs'.
This method has been recently applied to 57 bright EROs 
by Mannucci et al. (2002) using $J-\Ks$ vs $R-\Ks$.

\begin{figure*}
\epsscale{1.0}
\plotone{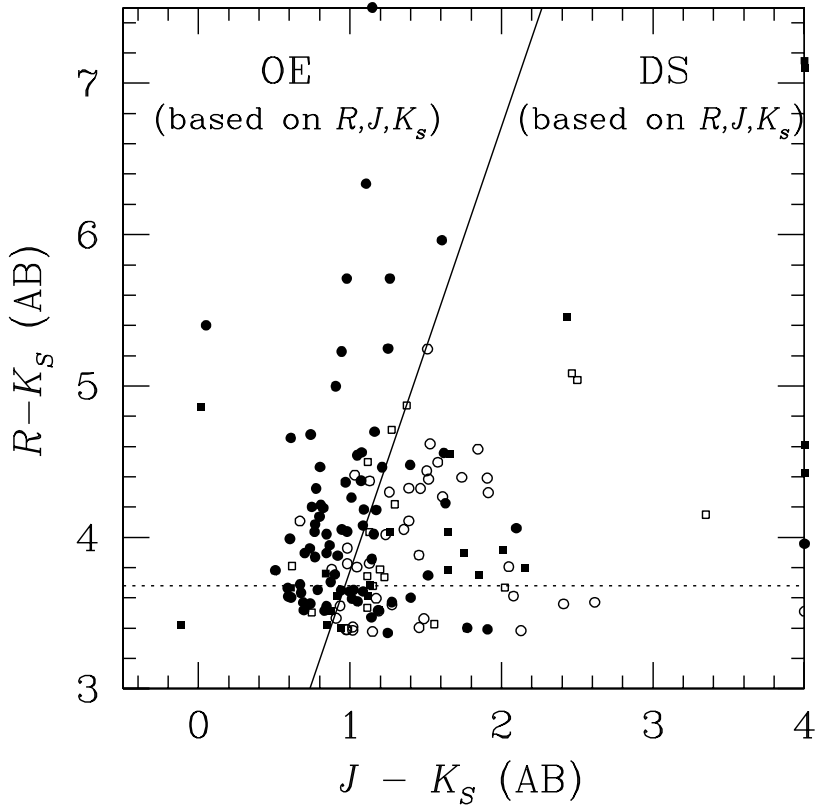}
\caption{$R-\Ks$ plotted against $J-\Ks$ for EROs in our sample 
         having $J$ magnitude.
         Filled symbols and open symbols indicate OEs and DSs, 
         respectively, classified by our spetrum fitting method.
         Objects at $1 \le z \le 2$ are shown by circles and 
         those at $z < 1$ or $z > 2$ are shown by squares.
         Objects with $J-\Ks > 3.5$ are plotted on 
         the $J-\Ks=4$ line, and those with $R-\Ks > 7.5$ 
         are plotted on the $R-\Ks=7.5$ line.
         The solid line correponds to the boundary for separation of
         OEs and DSs using $R-\Ks$ vs $J-\Ks$ 
         defined by Mannucci et al. (2002); 
         objects in the right-hand side of this line are classified
         as DSs.
         Mannucci et al.'s (2002) classification is valid 
         only for EROs satisfying $R-\Ks > 3.68$ 
         (horizontal dotted line) and $1 \le z \le 2$.
         It is expected that our method, which is based on 
         eight bandpasses, gives a better classification 
         over a wider range of redshift.
    \label{fig:JK_RK}}
\end{figure*}

As a check of the validity of our spectrum fitting method, 
we apply this $J-\Ks$ vs $R-\Ks$ method to our EROs.
Figure \ref{fig:JK_RK} shows the result.
Filled and open symbols indicate OEs and DSs, respectively, 
classified by our spectrum fitting method.
The solid line denotes the boundary for the two classes 
based on $J-\Ks$ vs $R-\Ks$ colors; objects redder in $J-\Ks$ 
than this line are classified as DSs.
According to Mannucci et al. (2002), their method is valid 
only for EROs satisfying $R-\Ks > 3.68$ 
(horizontal line in Figure \ref{fig:JK_RK}) 
and $1 \le z \le 2$, and cannot estimate redshift.

The agreement between the two methods is found to be satisfactory. 
Among the EROs with $R-\Ks > 3.68$, 
about 66 \% of the EROs classified as OEs by our method are also 
regarded as OEs by the $J-\Ks$ vs $R-\Ks$ method.
Similarly, 30 out of the 39 EROs which are classified as DSs 
by our method are located at the right-hand side of the solid line.
There are 23 EROs (with $R-\Ks > 3.68$) 
which are classified as OEs by our method
but are located at the right-hand side of the solid line.
However, 61 \% of them are at either $z<1$ or $z>2$, 
in which the $J-\Ks$ vs $R-\Ks$ method does not return 
a reliable answer.
In other words, if we restrict ourselves to those objects 
located at $1 \le z \le 2$,
82\% of OEs and 77\% of DSs classified by our method are 
found to be classified as OEs and DSs, respectively,
by the $J-\Ks$ vs $R-\Ks$ method.

Our spectrum fitting method uses eight bandpasses from $B$ to $\Ks$.
This wide-range and fine sampling of spectra enable us
to classify securely EROs over a wider range of redshift 
than covered by the simple $J-\Ks$ vs $R-\Ks$ method.
In addition, our method estimates photometric redshifts, 
which are essential for our studies of the evolution of OEs.

\subsection{Results of Classification}

\subsubsection{Relative Mix of OEs and DSs}

Out of the 247 EROs in our sample, 143 and 104 are classified 
as OEs and DSs, respectively, by our spectrum fitting method. 
The fraction
of OEs is thus 58 \%.
No significant difference is found in the distribution of 
$\chi^2$ values between the OE and DS populations.
Mannucci et al. (2002) have classified 57 EROs down to $K'=21.8$ 
by the $J-K$ vs $R-K$ method and found the fraction of OEs 
over the sum of OEs and DSs to be $21/42=50\%$. 
Cimatti et al. (2002) have taken spectra of 45 EROs with $\Ks<21$, 
identified about 70\% of them with OEs and DSs, 
and found that the relative fraction of OEs is $\simeq 50\%$.
These two estimates, both are based on spectra of EROs, 
are consistent with our result.

EROs have also been classified morphologically.
Stiavelli \& Treu (2001) have detected 30 faint EROs 
using $R-H_{160}$ color in an area imaged with HST/NICMOS. 
They have visually inspected HST/NICMOS images, 
to find that about 60\% of the 30 EROs belong to elliptical or S0 
galaxies.
Moriondo et al. (2000) have made a quantitative 
study of the morphology of 41 EROs found in images from 
the HST public archive. 
They have performed a fit of a generalized exponential profile 
($I \propto \exp((R/R_e)^n)$; S\`ersic 1982) to the images, 
and found that $50-80\%$ of the 41 EROs are elliptical-like 
objects.
Vanzella et al. (2001) have reported on the basis of 
a small sample of EROs in the Hubble Deep Field South 
that roughly a half of the EROs for which visual classification 
is successfully made look like elliptical galaxies.

Although morphology and spectrum are independent properties, 
a rough agreement on morphological and spectrum-based 
classifications has been reported (e.g., Vanzella et al. 2001). 
In other words, 
EROs which have morphological appearance of an elliptical galaxy 
tend to have a spectrum similar to those of passively evolving 
galaxies.

\subsubsection{Redshift Distributions}

\begin{figure*}
\epsscale{1.0}
\plotone{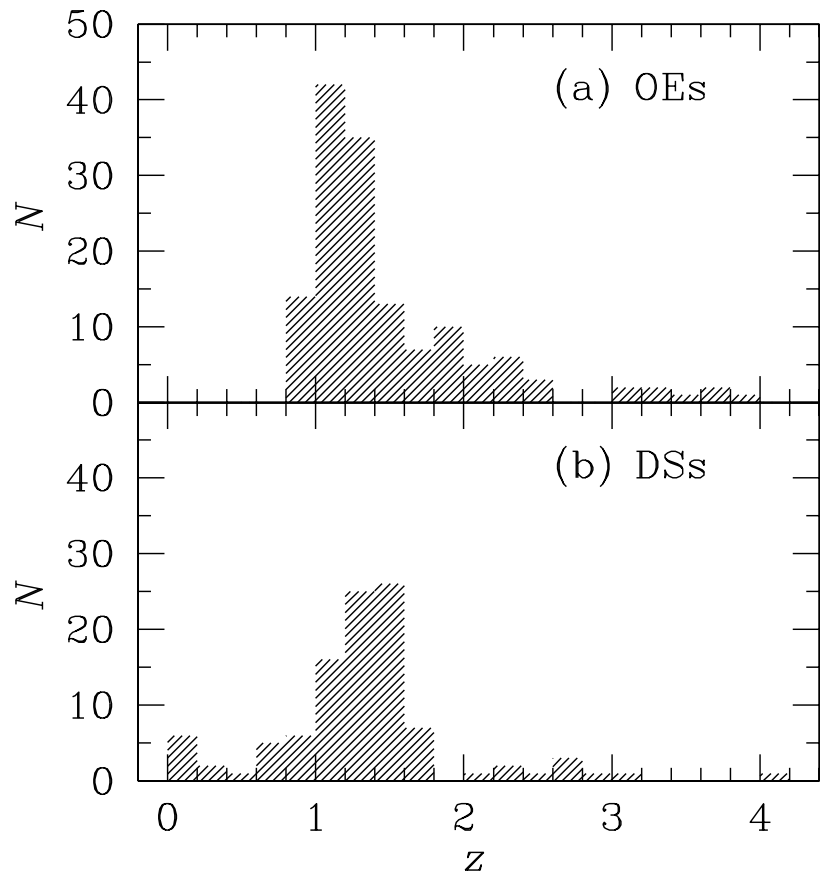}
\caption{Redshift distribution of all EROs in our sample.
         Panel (a): OEs. Panel (b): DSs.
    \label{fig:nzERO}}
\end{figure*}

We plot in Figure \ref{fig:nzERO} the redshift distribution 
of OEs (panel [a]) and DSs ([b]) in our sample. 
The redshift distribution of OEs in our sample has 
a peak at $z\sim 1$, with a tail toward higher redshifts. 
The majority of OEs are, however, located at $z<2$; 
only 14 \% of the total OEs have $z>2$.
It is worth noting that the distribution of our OEs has 
a sharp cut at $z=0.8$. 
We have defined the threshold for EROs, $R-\Ks \ge 3.35$, 
in order to select $z \ge 0.8$ objects. 
The sharp cut at $z=0.8$ suggests that the redshift 
estimation by the spectrum fitting method is reasonable.

The peak of the redshift distribution of DSs in our sample 
is at $z\sim 1.5$, somewhat higher than the OEs'.
Unlike the distribution of OEs, a tail is seen at $z<0.8$; 
this is probably because the adopted threshold ($R-\Ks \ge 3.35$) 
is based on passively evolving populations.

\begin{figure*}
\epsscale{1.0}
\plotone{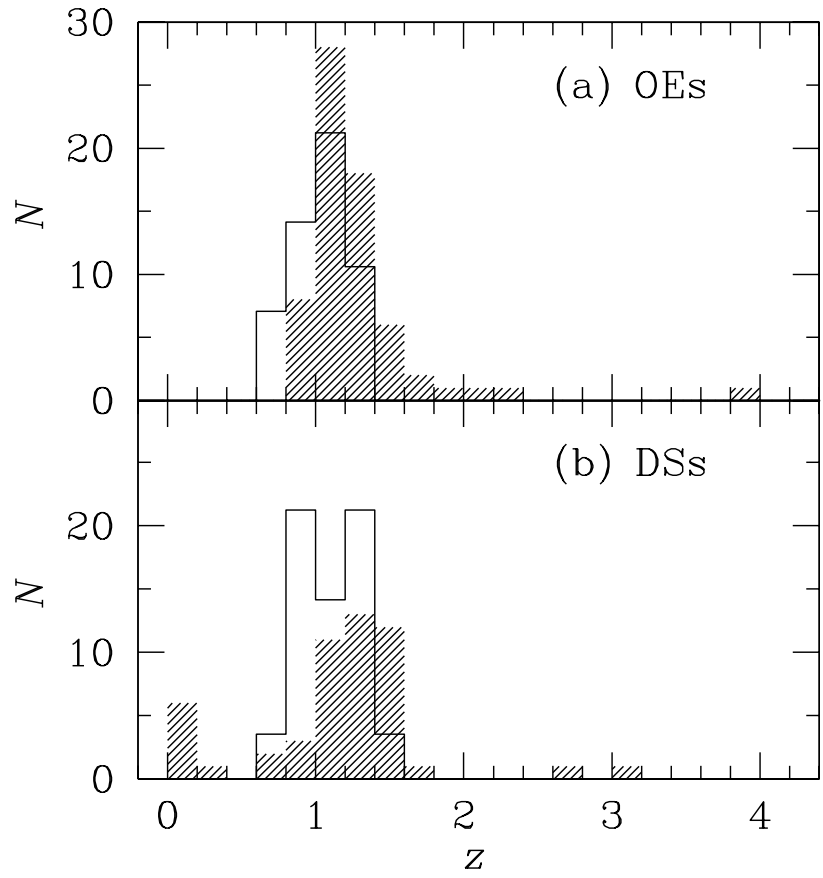}
\caption{Same as Fig. \ref{fig:nzERO}, but for EROs with $\Ks \le 21$.
         The open histograms show the data given in 
         Cimatti et al. (2002; K20 Survey).
    \label{fig:nzEROCimatti}}
\end{figure*}

Cimatti et al. (2002) have derived the redshift distribution 
of 14 OEs and 15 DSs for $\Ks<21$ from a spectroscopic survey of 
EROs with $R-\Ks \ge 5$ (Vega).
The completeness of their spectroscopic identification is $67\%$.
To compare their results with ours, 
we plot in Figure \ref{fig:nzEROCimatti} 
the redshift distribution of our 117 EROs 
brighter than $\Ks=21$ together with Cimatti et al.'s 
\footnote{Cimatti et al. present only the redshift distribution 
from a deeper ($\Ks<21.8$) spectroscopic survey. 
However, the cumulative completeness of this deeper 
survey is only $44\%$ and so the number of OEs and DSs 
identified at $21<\Ks<21.8$ is only one and three, respectively. 
Hence, we assume here that the redshift distribution given in 
Cimatti et al. is actually close to that for $\Ks<21$ objects.}.
Although relatively poor statistics of Cimatti et al.'s 
data do not allow a detailed comparison, 
a broad agreement is found in the redshift distribution 
between Cimatti et al.'s and our data for both OEs and DSs.
The median redshift of our sample is slightly higher than 
that of Cimatti et al.'s for both OEs and DSs. 
This may be partly due to the incompleteness of Cimatti et al.'s 
spectroscopy, since they claim that 
due to the noise at $\lambda > 9000$\AA, spectroscopic identification 
was difficult for OEs at $z>1.3$ and DSs at $z>1.5$.

\subsubsection{Bulge-to-Total Luminosity Ratios}

Table \ref{tab:BTdist} shows the distribution of $B$-band 
bulge-to-total luminosity ratio $B/T$ for 143 OEs in our sample.
All the OEs are found to have $B/T \ge 0.6$, 
among which 76 \% have $B/T > 0.9$.
The high average value of $B/T$ suggests that OEs in our sample 
are progenitors of the present-day elliptical galaxies.
On the other hand, the fraction of pure bulges with $B/T=1$ 
is not large ($24/143=17\%$) 
(Inclusion of objects with $B/T \ge 0.99$ increases 
the fraction to 34 \%).
These results indicate that the majority of the 
OEs selected in this study have a small but non-negligible 
amount of star forming component 
superposed on a dominant old population.

\section{RESULTS AND DISCUSSION}

\subsection{Number Counts}

\begin{figure*}
\epsscale{1.0}
\plotone{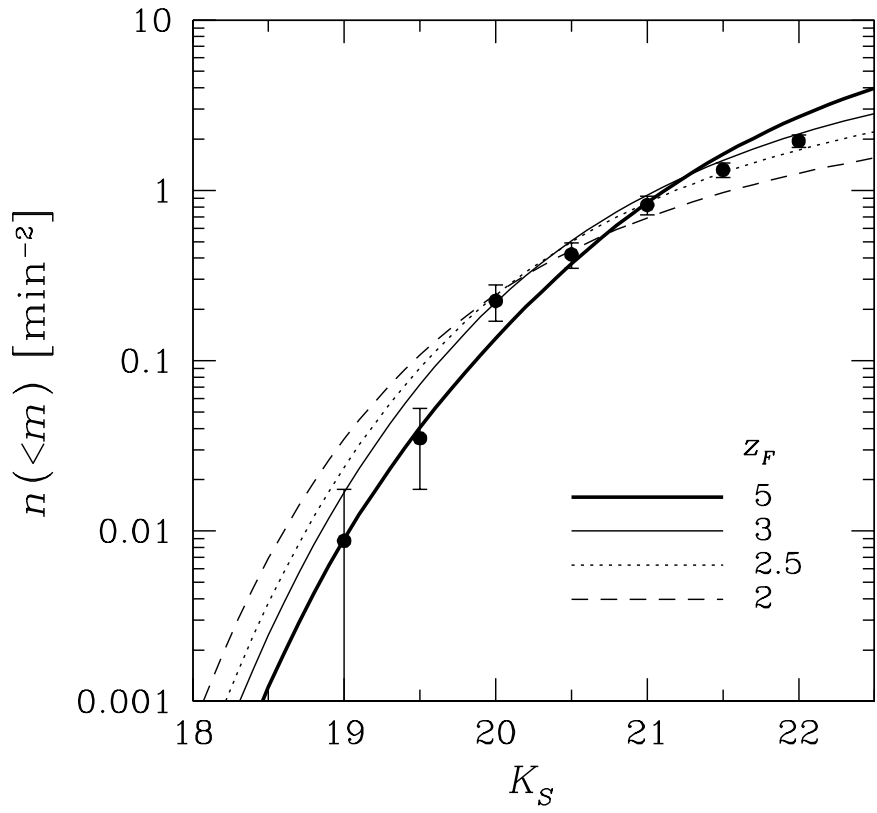}
\caption{Cumulative number counts of OEs in our sample.
         Predicted counts are plotted 
         by a thick solid line ($\zf=5$), 
         thin solid line ($3$), 
         dotted line ($2.5$), 
         and dashed line ($2$).
    \label{fig:nmOE}}
\end{figure*}

The cumulative number counts, $n(m)$, for OEs in our sample 
after the completeness and contamination corrections are plotted 
in Figure \ref{fig:nmOE}.
The slope of the counts 
is approximated as $n(m) \propto 10^{0.4 \times m}$ 
over $\Ks=21-22$, 
and no clear flattening is seen in the magnitude range observed.
The surface density of OEs with $\Ks<21$ 
is estimated to be $0.84 \pm 0.10$ arcmin$^{-2}$.
It is apparently a bit larger than Cimatti et al.'s (2002) 
estimation for OEs, $0.27-0.55$ arcmin$^{-2}$, 
which has not been corrected for completeness.
We compare the observed $n(m)$ of OEs with predictions by pure 
luminosity evolution models of elliptical galaxies.

\subsubsection{Prediction by Pure Luminosity Evolution Models}

We compute the number counts 
on the basis of the pure luminosity evolution (PLE)
model by KA97 described in \S 3, i.e., the model with $x=1.10$, 
$\tau_{\rm SF}=0.1$ Gyr, $\tau_{\rm infall}=0.1$ Gyr, 
$t_{\rm GW}=0.2$ Gyr. 
We examine four values for the formation redshift $\zf$: 
$\zf=5, 3, 2.5$, and 2.

The cumulative surface number density $n$ down to apparent $\Ks$ 
magnitude $m$ is computed as:
\begin{equation}
n = \int^{\zmax(\zf)}_{\zmin(\zf)} dz
    \int^{\Mf(\zf, m)}_{-\infty} dM
    \phi(M) {dV(z)\over{dz}},
\end{equation}
\noindent
where, $\zmin$ and $\zmax$ are the minimum and maximum redshifts 
at which the apparent color of the model galaxy satisfies 
the threshold for EROs, i.e., $R-\Ks \ge 3.35$, 
$\Mf$ is the absolute $\Ks$ magnitude 
corresponding to $m$, 
$\phi(M)$ is the $\Ks$-band luminosity function (LF) at $z=0$, 
and ${dV(z)\over{dz}}$ is the differential comoving volume.
$\Mf$ is defined as $\Mf = m - (m-M) - K(z) - E(\zf, z)$, 
where $(m-M)$ is the luminosity distance to an object at $z$ 
(distance modulus)
and $K(z)$ and $E(z)$ are the $K$ and $E$ corrections. 
Note that $\zmax$, $\zmin$, and $\Mf$ are dependent on $\zf$.

No local $\Ks$-band LF of pure elliptical galaxies 
is available in the literature.
Accordingly, we adopt the 
$B$-band LF for elliptical galaxies in the local universe derived by 
Marzke et al. (1994), which are characterized by the Schechter 
parameters, $\alpha=-0.85$, $\MB^\star = -19.37 - 5\log h$ 
(we adopt $B_{\rm AB}=B-0.14$ following Fukugita et al. 1995), 
and $\phi^\star = 0.0015h^3$ Mpc$^{-3}$. 
We convert $\MB^\star$ to $\MKs^\star$ using $\MKs = \MB - 2.23$ while
other parameters are unchanged.

There are a few other sources of the local $B$-band LF for 
E/S0 galaxies. 
Folkes et al. (1999) have obtained, using the 2dF Galaxy 
Redshift Survey data, 
$\alpha=-0.74\pm0.11$ and $\MB^\star = -19.75\pm 0.09 - 5\log h$ 
for local E/S0 galaxies; 
Marzke et al. (1998) have obtained $\alpha=-1.00\pm0.09$ 
and $\MB^\star = -19.51^{+0.10}_{-0.11} - 5\log h$ 
for E/S0 galaxies in the Second Southern Sky Redshift Survey 
data; 
Kochanek et al. (2001) have derived 
$\alpha=-0.92\pm0.10$ and $\MKs^\star = -23.53 - 5\log h$ 
(Vega) for E/S0 galaxies in the 2MASS galaxy data.
Accordingly, expected uncertainties in $\alpha$ and $\MKs^\star$ 
would be $\sim 0.2$ and $\sim 0.3$, respectively.
There are few measurements of $\phi^\star$ for ellipticals alone, 
but uncertainties in $\phi^\star$ could be as large 
as a factor of two, because the $\phi^\star$ values derived 
for E/S0s vary by the same amount 
(Marzke et al. 1998; Folkes et al. 1999; Kochanek et al. 2001).

\subsubsection{Comparison between Observations and Models}

Predicted counts are plotted in Figure \ref{fig:nmOE}
by a thick solid line ($\zf=5$), 
thin solid line ($3$), 
dotted line ($2.5$), 
and dashed line ($2$).  From a comparison of 
these predictions with the observed counts, 
we find that the $\zf=3$ and 2.5 models match 
the observed counts fairly well.
We assign large weights in the comparison on faint counts 
with small errors.
The $\zf=5$ model is found to predict the steepest slope of $n(m)$, 
and largely overshoot the observed counts at faint magnitudes 
although it predicts correct numbers at bright magnitudes.
On the other hand, the $\zf=2$ model underpredicts 
counts at faint magnitudes while it tends to produce 
a bit more galaxies than observed at bright magnitudes.
The $\zf=5$ and 2 models cannot be made consistent with 
the observation even if possible uncertainties in the local LF are 
taken into account.

This comparison shows that the range of $\zf$ 
for which models are consistent with 
the observed counts is rather limited, 
i.e., too large and too small $\zf$ values are not favored.
It should, however, be noted that 
the moderate matches of the $\zf=3$ and 2.5 models to 
the observed counts do not necessarily 
mean that OEs obey these models at any redshifts.
We will discuss this in the next subsection 
on the basis of the luminosity function.

Daddi et al. (2000b) have compared surface densities of 
EROs at $\Ks < 21$ 
selected by $R-\Ks$ colors from an area of 850 arcmin$^2$ 
with predictions of passive evolution models, 
to find a good agreement between the observed surface densities 
of EROs and those predicted by passive evolution models 
of $\zf \gsim 2.5$. 
They have not classified EROs into OEs and DSs, 
but they claim that their results do not change significantly 
if the fraction of DSs in their sample is less than $30\%$.
In this sense, their results appear to be roughly consistent with ours 
since the fraction of DSs in our sample, $\simeq 40\%$, 
is close to $30\%$.

\subsection{Luminosity Functions}

We construct the rest-frame $B$-band LF for OEs in our sample 
for two redshift bins, $1 \le z \le 1.5$ and $1.5 \le z \le 2.5$, 
using the simple $1/\Vmax$ method.
We do not derive LFs at $z>2.5$ because the OEs 
in this range are too few.
For each OE, the rest-frame $B$ magnitude is computed 
from the best-fit spectrum.

\begin{figure*}
\epsscale{1.0}
\plotone{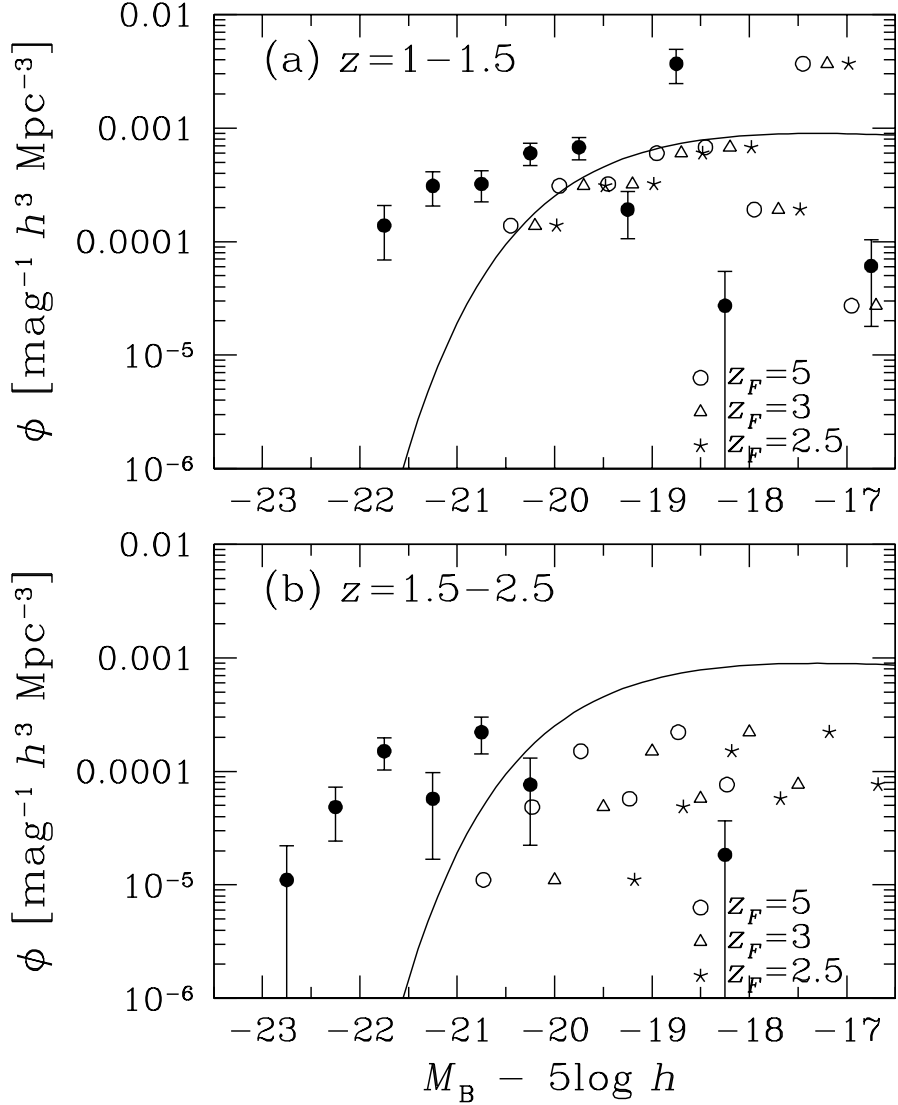}
\caption{Rest-frame $B$-band luminosity functions (LFs) 
         of OEs (filled circles). 
         Panel (a): objects at $z=1-1.5$.
         Panel (b): objects at $z=1.5-2.5$.
         The solid line denotes the local $B$-band luminosity 
         function for ellipticals given in Marzke et al. (1994).
         The open circles, open triangles, 
         and stars indicate predictions 
         at $z=0$ which are calculated by dimming the observed 
         LF to the present on the basis of the passive evolution 
         models with $\zf=5$, $3$, and $2.5$, respectively, 
         adopted in \S 5.1.
    \label{fig:LFOE}}
\end{figure*}

Figure \ref{fig:LFOE} plots the LFs derived in this way 
(filled circles with error bars).
The solid lines in Figure \ref{fig:LFOE} denote 
the local $B$-band luminosity function for ellipticals 
(Marzke et al. 1994) 
adopted in \S 5.1 to predict number counts.

A comparison of the LF derived for the range $z=1-1.5$ with 
the local measurement finds that while the number density of 
fainter ($M \gsim -19$) OEs at $z=1-1.5$ is comparable 
to that of the local ellipticals, OEs with $M \lsim -20$ 
are much more numerous than in the local universe.
Roughly speaking, the LF at $z=1-1.5$ can be fit by the local LF 
if a dimming of $M^\star$ by $\sim 1.3$ is assumed 
from $z=1.25$ to the present epoch.
As we see later, this amount of dimming is predicted 
by the PLE model with $\zf\ge 3$.
Similarly, the number density of $z=1.5-2.5$ OEs with 
$M \lsim -21$ is found to be much higher than that observed 
in the local universe.
However, unlike the LF at $z=1-1.5$, 
it is difficult to fit the LF at $z=1.5-2.5$ to the local LF 
by a shift of $M^\star$ alone, 
because the amplitude of the $z=1.5-2.5$ LF is lower than 
that of the local LF.

The open circles indicate predictions 
at $z=0$ which are calculated by dimming the observed 
LFs at $z=1-1.5$ (panel [a]) and at $z=1.5-2.5$ (panel [b]) 
to the present epoch on the basis of 
the passive evolution model with $\zf=5$ adopted in \S 5.1.
Similarly, the open triangles and stars are predictions 
from the $\zf=3$ and 2.5 models, respectively.
If all ellipticals seen at present obey the passive evolution 
of, say, $\zf=5$, 
then in both panels of Figure \ref{fig:LFOE} 
the open circles should be on the solid line.
In panel (a), the open circles and open triangles match the local LF 
over almost the whole magnitude range.
Taking a smaller value for $\zf$ makes the amount of dimming 
larger, and thus the LFs evolved to $z=0$ will become fainter 
than that of the observed local LF.
For instance, the dimming predicted by the $\zf=2.5$ model appears 
to be too large by $\sim 0.5$ mag.

In Figure \ref{fig:LFOE} (b), 
the difference between the observed local LF 
and the prediction from the $z=1.5-2.5$ LF 
is found to be very large for all the three models. 
Even for the $\zf=5$ model, 
which predicts the smallest amount of dimming among the three, 
the amplitude of the predicted LF is lower than that of 
the observed local LF by a factor of $\sim 3$ or so 
at any magnitudes.
If the $\zf=3$ and 2.5 models are taken, discrepancies at 
bright magnitudes become quite large; 
basically no galaxies brighter than $-20$ mag are predicted 
by these models.
This result means that 
the fraction of the progenitors of the present-day ellipticals 
which have obeyed passive evolution at $z>1.5$ is 
significantly low. 
We have found in \S 5.1 that the $\zf=2.5$ and 3 models 
match the observed number counts of OEs.
The discrepancies found in the LF at $z=1.5-2.5$ implies 
that the matches seen in the number counts are superficial. 
Simply speaking, the number density of OEs predicted by 
these models is biased towards $z>1.5$ 
while the total number of OEs integrated over redshift 
by these models roughly matches the observed counts.

We have not derived LFs for $z>2.5$, but a rough estimate 
shows that the number of observed OEs is much smaller 
than the prediction of the $\zf=5$ model.
Note that models with $\zf\le 3$ predict no 'OEs' at $z>2.5$ 
since galaxies obeying these models 
do not satisfy $R-\Ks\ge 3.35$ at beyond $z=2.5$. 

\subsection{Colors}

\begin{figure*}
\epsscale{1.0}
\plotone{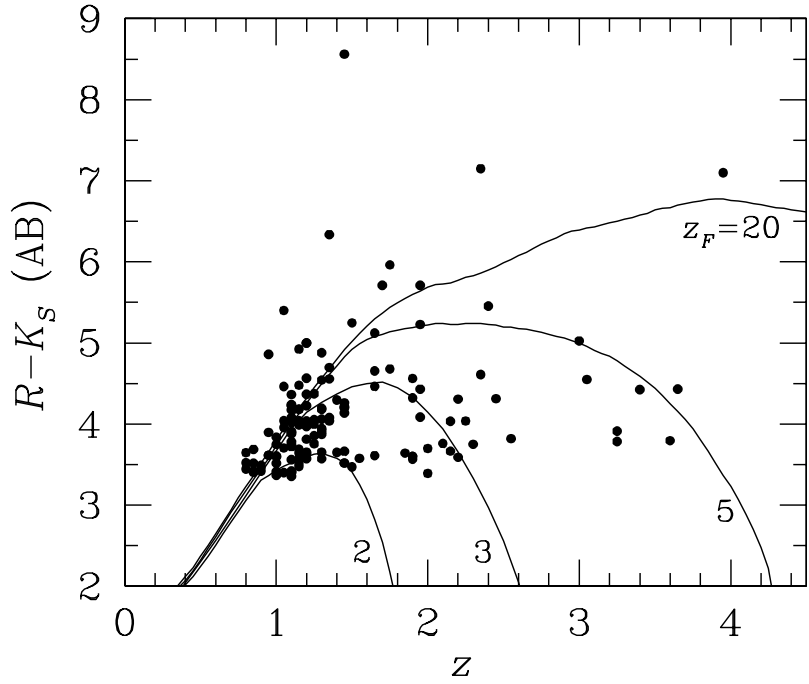}
\caption{$R-\Ks$ of OEs plotted against redshift.
         The four solid lines indicate predictions by passive 
         evolution models with $\zf=20, 5$, 3, and 2.
    \label{fig:RK_z}}
\end{figure*}

Figure \ref{fig:RK_z} plots observed $R-\Ks$ color as a function of 
redshift for OEs in our sample.
The four solid lines indicate predictions by passive evolution 
models with $\zf=20, 5$, 3, and 2.
It is found that the number of OEs having colors 
clearly redder than the prediction by the $\zf=5$ model 
is very small.
The observed OEs have a wide range of color 
at any redshift, meaning that the OEs cannot 
be reproduced by a single, passive evolution model 
of a specific $\zf$.

\begin{figure*}
\epsscale{1.0}
\plotone{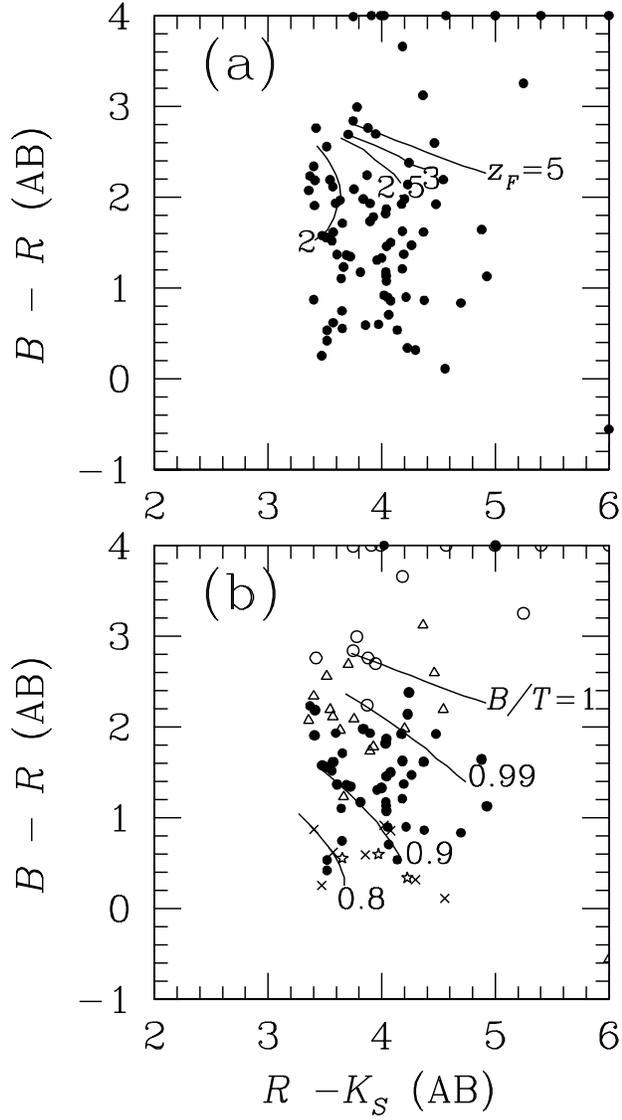}
\caption{$R-\Ks$ plotted against $B-R$ for OEs at $z=1-1.5$.
         Objects with $R-\Ks > 6$ are plotted on 
         the $R-\Ks=6$ line, and those with $B-R > 4$ 
         are plotted on the $B-R=4$ line.
         Panel (a): Lines show predictions from passive 
         evolution models with $\zf=5, 3, 2.5$, and 2.
         Panel (b): Lines show predictions from bulge plus disk
         models with $B/T=1, 0.99, 0.9$, and 0.8 ($\zf=5$).
         For the data, different symbols indicate different
         $B/T$ values; 
         open circles : $B/T=1$,
         open triangles : $0.99 \le B/T < 1$,
         filled circles : $0.9 \le B/T < 0.99$,
         crosses : $0.8 \le B/T < 0.9$,
         and stars : $B/T < 0.8$.
    \label{fig:RK_BR}}
\end{figure*}

Figure \ref{fig:RK_BR}
plots $R-\Ks$ against $B-R$ for OEs at $z=1-1.5$.
The four lines in panel (a) indicate evolutionary 
tracks from $z=1.5$ to 1 
of passive evolution models with $\zf=5, 3, 2.5$, and 2.
The majority of the OEs are found to be located 
outside the region between the $\zf=5$ and 2 tracks.
This suggests that most of the OEs do not have a color 
consistent with passive evolution models of $\zf \ge 2$.
Panel (b) of Figure \ref{fig:RK_BR} 
plots the same data as for panel (a), 
but the four lines indicate colors from $z=1.5$ to 1 
of a passively evolving bulge of $\zf=5$ on which a disk 
component is superposed with the rest-frame $B/T$ ratio 
of 1 (i.e., pure bulge), 0.99, 0.9, and 0.8.
For the data, symbols are changed according to $B/T$ of 
the best-fit spectrum; 
open circles for $B/T=1$, open triangles for $0.99 \le B/T < 1$, 
filled circles for $0.9 \le B/T < 0.99$,
crosses for $0.8 \le B/T < 0.9$, 
and stars for $B/T < 0.8$.
It is found that OEs with lower $B/T$ have bluer $B-R$ color, 
with $R-K$ color almost unchanged. 
It is also found that the majority of the OEs fall between 
the $B/T=1$ and 0.8 lines.
We do not intend to say that most OEs are on evolutionary 
tracks of a bulge $+$ disk galaxy of $\zf=5$, 
because the ages of the best-fit spectra span over a wide range.
The point here is that the observed OEs are expressed 
by bulge $+$ disk models much better than by pure bulges 
of passive evolution of different ages.
This implies that the majority of the OEs in our sample
have non-negligible star formation.
If we calculate the star formation rate from the best-fit 
spectrum, we obtain $0.9 h^{-2} M_\odot$ yr$^{-1}$ 
for an OE with $\Ks=21$ and $B/T=0.9$ at $z=1.25$.

\subsection{Clustering}

\begin{figure*}
\epsscale{1.0}
\plotone{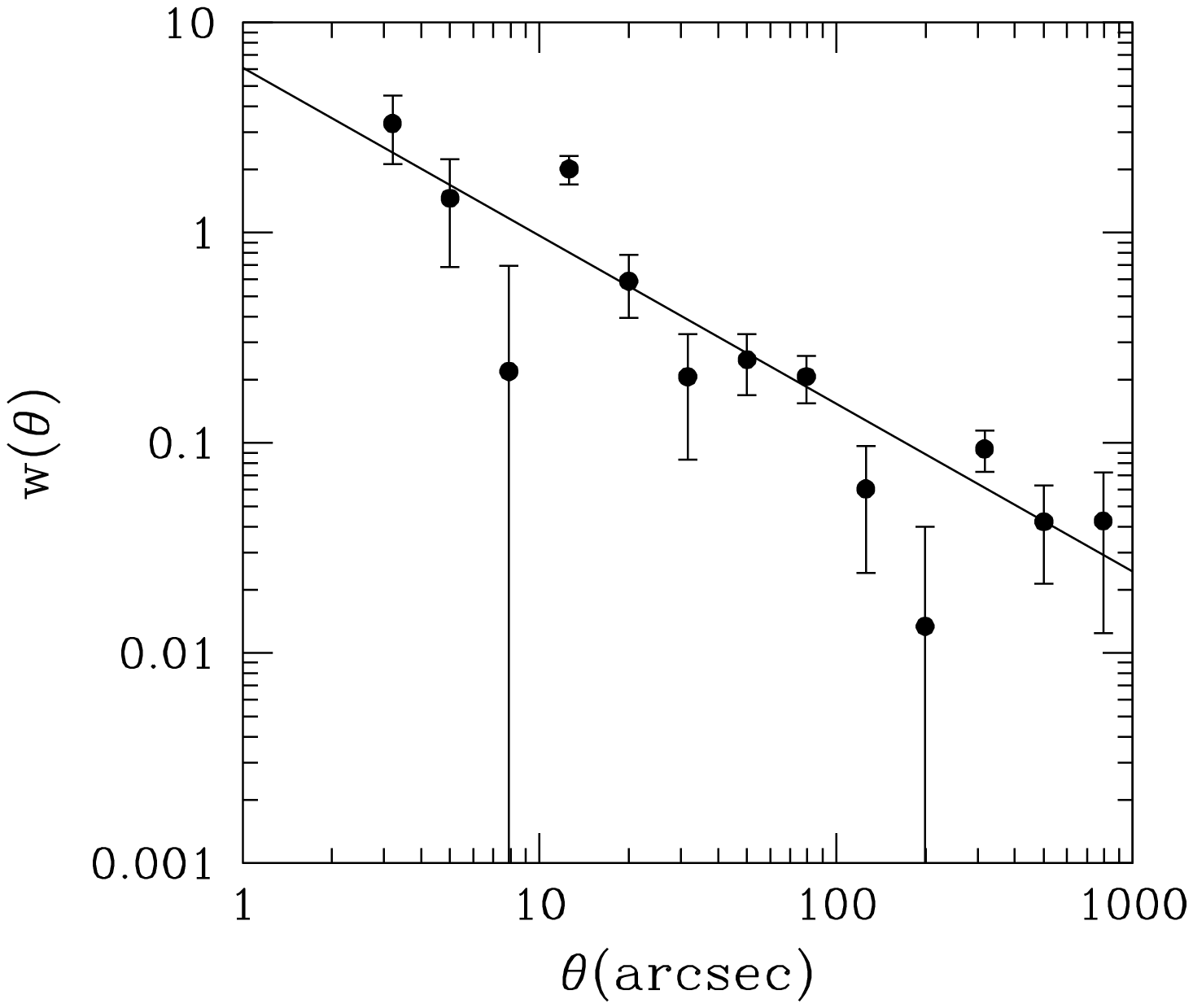}
\caption{Angular correlation function of OEs.
    The solid line shows the best fit power law with 
    $\omega(\theta)=A_\omega \theta^{-0.8}$ to the data
    over the range of $3'' \le \theta \le 150''$.
    \label{fig:ACFoe}}
\end{figure*}

We examine clustering of OEs in our sample.
We derive the angular two-point correlation function, 
$\omega$($\theta$), 
using the estimator defined by Landy \& Szalay (1993),
$
\omega_{obs}(\theta)
  = [DD(\theta)-2DR(\theta)+RR(\theta)]/RR(\theta),
$
where $DD(\theta)$, $DR(\theta)$, and $RR(\theta)$ are numbers of
galaxy-galaxy, galaxy-random, and random-random pairs normalized by
the total number of pairs in each of the three samples.
The real correlation function $\omega(\theta)$ is
offset by the integral constant ($IC$: Groth \& Peebles 1977);
$\omega(\theta)= \omega_{obs}(\theta)+IC$.
The value of $IC/A_\omega$ in our sample is calculated to be 0.0124, 
where $A_\omega$ is the amplitude of the true angular correlation 
function at $1''$ (see below).
The resulting angular correlation function 
is plotted in Figure \ref{fig:ACFoe} after application of $IC$.
This figure 
shows that OEs are clearly clustered on the sky.
The amplitude of the angular correlation function increases 
with decreasing $\theta$, and takes $\gsim 1$ at 
$\theta \lsim 10''$.

We fit a power law, 
$\omega(\theta)=A_\omega (\theta^{-0.8} - IC)$, to the data points 
over the range of $3'' \le \theta \le 150''$, 
to find $A_\omega = 6.1 \pm 0.9$ arcsec$^{0.8}$, 
or equivalently $(8.7 \pm 1.3) 10^{-3}$ degree$^{0.8}$.
This best-fit power-law is shown by a solid line 
in Figure \ref{fig:ACFoe}.
The power law with an index of $-0.8$ is found 
to approximate the data well.

Several papers have measured $A_\omega$ of EROs, 
but all of them are not for the OE population 
but for the ERO population as a whole. 
One of the measurements is given in Daddi et al. (2000b), 
who select 281 EROs of $R-\Ks \ge 5$ (Vega) down to $\Ks=19.2$ (Vega) 
in a 447.5 arcmin$^2$ area and find 
$A_\omega = (13 \pm 1.5) 10^{-3}$ degree$^{0.8}$.
Their selection threshold for EROs is the same as in this 
paper, $R-\Ks \ge 3.35$ (AB).
In order to compare with their result, 
we measure $A_\omega$ for all EROs with $\Ks \le 21$ 
(corresponding to $\Ks \le 19.2$ in the Vega system) in our sample. 
We find $A_\omega = (11 \pm 2) 10^{-3}$ degree$^{0.8}$, 
in good agreement with Daddi et al.'s measurement.

Using the Limber transformation 
(e.g., Peebles 1980, Efstathiou et al. 1991) 
and adopting the distribution of photometric redshifts of our OEs, 
we estimate from $A_\omega$ the spatial correlation length 
$r_0$ for OEs to be $11 \pm 1 h^{-1}$ Mpc, 
where we assume no evolution in comoving space 
\footnote{The redshift distribution of OEs has a long high-redshift 
tail, which may be partly due to errors in photometric redshifts.
However, OEs contained in the tail do not have 
a large contribution to the estimation of $r_0$.
Indeed, the $r_0$ value does not change significantly 
even if we restrict the redshift range to $z=0.8-1.6$, 
in which 73\% of the all OEs are included.}. 

Norberg et al. (2002) have measured 
$r_0$ for early-type galaxies with $\MB =-19.5$ - $-20.5$ 
and $\MB =-21$ - $-22$ in the local universe to be 
$\approx 6 h^{-1}$ Mpc and $\approx 10 h^{-1}$ Mpc, respectively.
Guzzo et al. (1997) have obtained 
$r_0 = 8.35^{+0.75}_{-0.76} h^{-1}$ Mpc 
for early-type galaxies with $\MB \ge -19.5$ 
in the Pisces-Perseus supercluster region.
Willmer, da Costa, \& Pellegrini (1998) 
have obtained $r_0=5.7 \pm 0.8 h^{-1}$ Mpc 
for early-type galaxies with $\MB \le -19.5$ 
in field regions of the local universe.
Since the typical luminosity of OEs in our sample is 
$\MB \sim -20$, the clustering of the OEs in our sample 
is found to be stronger than (or at least as stong as) those 
of the present-day early-type galaxies with similar luminosities.
This may imply that OEs found at $z>0.8$ are selectively 
located in regions which will become clusters or groups 
of galaxies, where most of the present-day early-type galaxies 
are found.

Our measurement of $r_0$ for OEs is consistent with
that for the OE population at $z\sim 1$ 
estimated by Daddi et al. (2002), $r_0 \simeq 5.5 - 16 h^{-1}$ Mpc,
based on a small ($N=15$) but spectroscopically classified sample 
of OEs.
Our measurement is also similar to those for the ERO population 
as a whole;
McCarthy et al. (2001) have selected red ($I-H \ge 3$ in Vega) 
galaxies in the Las Campanas Infrared Survey and have estimated 
$r_0$ of these galaxies to be $\approx 9-10 h^{-1}$ Mpc 
at $z \simeq 1$ based on photometric redshifts.
Daddi et al. (2001) have obtained $r_0 = 12 \pm 3 h^{-1}$ Mpc
for EROs with $R-\Ks \ge 5$ (Vega) brighter than $\Ks = 19.2$ 
(Vega) assuming that the redshift distribution for 
EROs is described by a passive evolution model.

Kauffmann et al. (1999) have calculated 
$r_0$ of early-type galaxies for $z=0-2.5$ 
in a $\Lambda$ Cold Dark Matter model.
They have found that $r_0$ remains almost constant over this 
redshift range, $r_0 \approx 7 h^{-1}$ Mpc.
Although this value is apparently smaller 
than that found in this study, 
their selection criteria for early-type galaxies are different 
from ours.
Thus, a detailed comparison requires corrections 
for these differences, which is beyond the scope of this paper.

\subsection{Implications for the Evolution of Ellipticals at $z>1$}

We have found that there exists a population of galaxies 
(which we call OEs) at $z>0.8$ whose red $R-\Ks$ colors are 
consistent with predictions 
of passive evolution models for elliptical galaxies.
The strong clustering ($r_0 \sim 10 h^{-1}$ Mpc) 
found for these OE galaxies supports the assumption that 
they are high-redshift counterparts of the present-day 
elliptical galaxies.
The spatial distribution of OEs appears 
to be strongly biased against the distribution of dark matter, 
since $r_0$ of the dark matter, which is a decreasing function 
of redshift, should be much smaller than $\sim 10 h^{-1}$ Mpc 
at $z>0.8$.

We have found that the spread in color of our OEs 
is considerably large at any redshift; 
only a small fraction of OEs are redder than predictions 
of a pure passive evolution model with $\zf=5$.
This implies that ellipticals at $z>0.8$ are heterogeneous
in terms of the stellar population.
In other words, colors of OEs at $z>0.8$ cannot 
be reproduced by any single passive evolution model with a fixed $\zf$.
Similar results have been obtained on colors 
of morphologically selected E/S0s at high redshifts 
on the basis of small samples from deep but narrow field surveys 
(e.g., Franceschini et al. 1998; Kodama et al. 1999; 
Rodighiero et al. 2001).
It is also found that $R-\Ks$ vs $B-R$ colors of the OEs 
in our sample are reproduced much better by bulge $+$ disk spectra 
than by spectra of simple passively evolving bulges, 
suggesting the presence of a non-negligible star formation activity.

Many authors have claimed a possible decrease in the number density 
of red galaxies at $z \gsim 1$ on the basis of 
observed surface densities on the sky.
For example, Barger et al. (1999) have found 
in a 61.8 arcmin$^2$ area to a depth equivalent to $K=20.1$ 
(Vega) that only a small population of galaxies have 
colors redder than $I-K=4$ (Vega),
the color expected for an evolved elliptical galaxy at $z>1$ 
(See also McCracken et al. 2000).
In this study, we have constructed LFs of OEs 
to locate the era when the number density of OEs changed 
and to discuss the number evolution of ellipticals.
We have found that the spatial number density of OEs 
at $z \ge 1.5$ is lower than that of the present-day ellipticals 
by a factor of $\sim 3$ (or more) if a dimming of brightness due to 
passive evolution is taken into account, 
while the LF of OEs at $z=1-1.5$ matches 
the local LF of ellipticals if a dimming of 1.3 mag from $z=1.25$ 
to the present epoch (the amount of dimming expected for 
the passive evolution model of $\zf\ge3$) is assumed. 

Our finding on the number evolution for OEs 
is in parallel with the result obtained by Rodighiero et al. (2001) 
for morphologically selected E/S0s.
Rodighiero et al. (2001) have examined the distribution 
of (photometric) redshifts for 69 morphologically selected E/S0 
galaxies at $K \le 20.15$ (Vega), 
among which 38 are at $z>0.8$, 
in a total of 11 arcmin$^2$ area of the HDFN, HDFS, and HDFS/NICMOS. 
They have found that massive E/S0s
tend to disappear from flux-limited samples at $z>1.4$ 
(see also Franceschini et al. 1998, Nakata et al. 1999).
They have reported that the number of E/S0s detected at $z>1.5$ 
is about five times lower than predictions from passive evolution 
models with $\zf \ge 3$, 
though it is based on poor statistics 
(they detected only two galaxies at $z>1.5$).

These findings collectively suggest that 
a significant fraction of ellipticals seen at present 
do not have counterparts at $z \gsim 1.5$ 
in terms of either old stellar population or morphology.
We discuss two possibilities which could explain the lack of 
$z>1.5$ ellipticals.
One is that most ellipticals obey pure luminosity evolution 
(i.e., no merging) even at these redshifts 
but that residual star formation superposed on old 
populations is so strong that they drop 
out of our color selection for EROs;
galaxies with non-negligible star formation 
are likely to have morphologies significantly 
different from smooth $1/4$-law profiles.
Indeed, the number of OEs in our sample which are fit by a spectrum 
of pure passive evolution (i.e., bulge) is very limited, 
and about 24 \% of our OEs are fit by bulge $+$ disk models 
with $B/T \le 0.9$. 
Our threshold $R-\Ks=3.35$ selects galaxies with $B/T \gsim 0.8$
at $z=2$ ($\zf=5$).
Thus, if ellipticals at $z=2$ have star formation equivalent to 
$B/T < 0.8$, they will not enter our sample.
Note that at $z=1.25$ 
the $R-\Ks=3.35$ threshold corresponds to $B/T \gsim 0.7$,
i.e., the minimum value of $B/T$ passing 
the threshold increases with redshift. 
This effect may also contribute to the observed decrease 
in the number density of OEs from $z=1-1.5$ to $z=1.5-2.5$, 
but we expect that this effect is relatively small 
as long as the distribution of $B/T$ for OEs does not change 
significantly from $z=1-1.5$ to $z=1.5-2.5$, 
since the fraction of OEs with $B/T \le 0.8$ is only 12\% 
at $z=1-1.5$.
Thus, if we want to attribute the decrease in number density 
of OEs at $z \ge 1.5$ to residual star formation, 
we must assume that majority of OEs seen at $z=1-1.5$ 
had $B/T$ smaller than 0.8 at $z \ge 1.5$.

The other possibility is that the present-day ellipticals 
have been formed from smaller fragmentary galaxies whose 
morphologies are not necessarily elliptical.
Interestingly, Rodighiero et al. (2000) have found 
the lack of massive, late-type galaxies at $z > 1.4$ 
in their sample from the HDFN.
Galaxies of any morphological types might have been 
significantly less massive (i.e., fainter) at $z>1.5$ 
than predicted from pure luminosity evolution models.
We cannot prove or disprove this possibility using our data, 
but an interesting trend is noted that 
bright ($\le -20$) OEs predicted at $z=0$ 
from the $z=1.5-2.5$ LF using the $\zf=3$ and 2.5 models
is much less numerous than the observed local ellipticals 
with the same luminosity.

We are planning to estimate photometric redshifts 
for all galaxies in our $\Ks$-limited sample, 
to derive color-dependent LFs for them.
These LFs will give us strong constraints on the number 
and luminosity evolutions of galaxy population as a whole 
with high statistical significance. 

\section{PROPERTIES OF DUSTY STARBURST GALAXIES}

In this section, we briefly summarize the properties of the dusty 
starburst galaxies found in our data.
Out of the 247 EROs in our $\Ks$-limited sample, 104 are found 
to be DSs according to the spectrum fitting.

As seen in \S 4, the DSs are distributed over a wide range of 
redshift, though their abundance relative to OEs becomes lower 
at $z \gsim 2$.
A small fraction of them are located at $z<0.8$.

\begin{figure*}
\epsscale{1.0}
\plotone{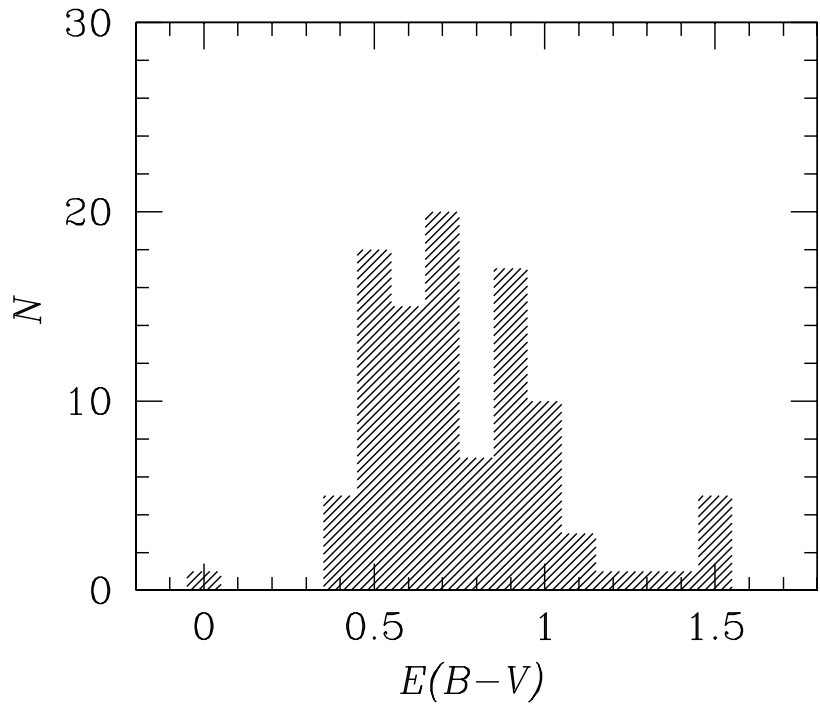}
\caption{Distribution of $E(B-V)$ for DSs in our sample.
    \label{fig:EBV}}
\end{figure*}

Figure \ref{fig:EBV} shows the distribution of $E(B-V)$ 
for DSs in our sample.
The DSs are found to have $E(B-V) \sim 0.4-1.2$, 
with a median value of $\simeq 0.8$. 
This corresponds to $A_V \simeq 2.5$ mag.
The $E(B-V)$ distribution of our DSs is 
in good agreement with that of Cimatti et al.'s (2002).
They have found that the average spectrum 
of DSs for which they made spectroscopy 
is fit by a local starburst galaxy with $E(B-V) \sim 0.8$.
They have also found that the global shape of the continuum 
and the average $R-\Ks$ color of their DSs can be 
reproduced by synthetic spectra of star forming galaxies 
of $x=1.35$ with $0.6 < E(B-V) < 1.1$ 
(they have also adopted Calzetti's [1997] extinction law).

We estimate the star formation rate of DSs in our sample 
using the luminosity at 2800\AA, $L_{2800}$,  
adopting the formula given in 
Madau, Pozzetti, \& Dickinson (1998): 
$SFR [M_\odot \hspace{3pt}{\rm yr}^{-1}]
= L_{2800} [{\rm erg}\hspace{3pt}{\rm s}^{-1} {\rm Hz}^{-1}] 
/ (7.9 \times 10^{27})$.
We find that the dust-uncorrected $SFR$ spans over 
$10^{-1}-10^1$ $M_\odot \hspace{3pt}{\rm yr}^{-1}$, 
with a median of 1.2 $M_\odot \hspace{3pt}{\rm yr}^{-1}$.
If we correct for the dust extinction, the median value 
increases by two orders of magnitude,
which is close to those expected for initial starburst phases 
of elliptical galaxies in passive evolution models.
If such high star formation rates last for longer than 
$10^8$ yr or so, 
these DSs will be recognized as elliptical galaxies 
in late epochs, and thus 
part of the ellipticals seen at present could be
descendants of these DSs.
This may be able to partly account for 
the observed lack of OEs at high redshift in comparison 
with the present-day ellipticals. 

\begin{figure*}
\epsscale{1.0}
\plotone{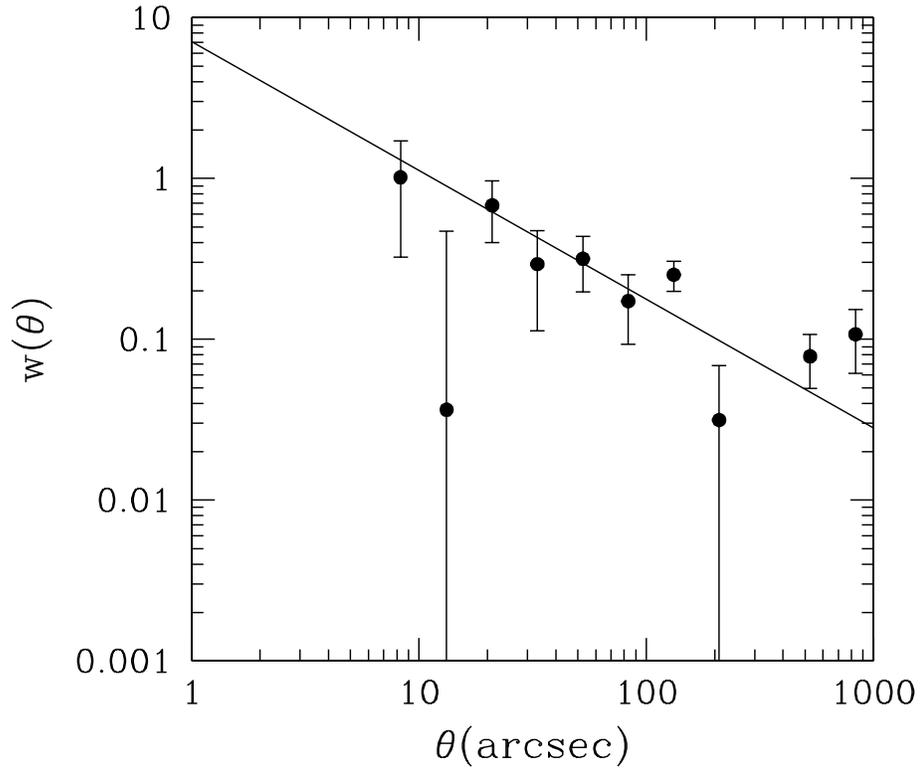}
\caption{Angular correlation function of DSs.
    The solid line shows the best fit power law with 
    $\omega(\theta)=A_\omega \theta^{-0.8}$ to the data 
    over the range of $20'' \le \theta \le 150''$.
    \label{fig:ACFds}}
\end{figure*}

Finally, we derive the angular correlation function 
for DSs in the same manner as for OEs.
The result is plotted in Figure \ref{fig:ACFds}.
The amplitude of angular clustering of DSs is 
found to be similar to that of OEs at $\theta \gsim 20''$, 
but is much lower at smaller scales.
A fit of $\omega(\theta)=A_\omega (\theta^{-0.8} - IC)$ 
to the data points over $20'' \le \theta \le 150''$ 
gives $A_\omega = 7.1 \pm 1.8$, 
corresponding to $r_0 = 12 \pm 2 h^{-1}$ Mpc. 
If taken at face value, this $r_0$ value is very similar 
to that for OEs. 

If the clustering of DSs is really as strong as that of OEs, 
this would suggest that, similar to OEs, 
DSs are also formed in high density peaks of the dark matter 
fluctuations which will become present-day clusters or groups. 
This appears to support the hypothesis shown above that DSs are 
progenitors of part of the present day E/S0s.
Daddi et al. (2002) have reported that $r_0$ of DSs 
at $z\sim 1$ is less than $\sim 2.5 h^{-1}$ Mpc, 
which is apparently inconsistent with our measurement.
However, their estimation is derived from only 18 DSs found
in 50 arcmin$^2$.
On the other hand, their classification is based on 
spectral features of objects, a more direct indicator 
than we use in this study.
Our DS sample could suffer from a non-negligible contamination 
by OEs due to mis-classification.
In addition, Figure \ref{fig:ACFds} shows 
that the clustering of DSs on $\lsim 10''$
may be much weaker than an extrapolation of 
the best-fit $A_\omega \theta^{-0.8}$ determined at 
$20'' \le \theta \le 150''$.
An accurate measurement on small scales using a larger sample 
may reduce $A_\omega$ and thus $r_0$ significantly.

\section{SUMMARY}

In order to study properties of elliptical galaxies at $z>0.8$, 
we have constructed a large sample of Extremely Red Objects 
(EROs) from $BVRi'z'JH\Ks$ multicolor data 
of a 114 arcmin$^2$ area in the Subaru/XMM Deep Survey Field 
taken with the Subaru Telescope ($B$ to $z'$) 
and the UH 2.2m Telescope ($JH\Ks$).
We have detected 247 EROs with $R-\Ks \ge 3.35$ down to $\Ks = 22.1$ 
in the data.
This is the largest multicolor sample of EROs 
covering $B$ to $\Ks$ bands ever obtained.

By fitting template spectra of old ellipticals (OEs) and 
young, dusty starbursts (DSs) at $z=0-4$ to the multicolor data, 
we have classified the EROs into these two classes 
and estimated their redshifts.
In order to express modest star formation superposed on 
old stellar population, we include in the OE templates 
a wide range of combination of an old population (bulge component) 
and a young population.
We have compared our spectrum fitting method with 
a simple classification method based on $J-\Ks$ vs $R-\Ks$ colors, 
which was introduced by Pozzetti \& Mannucci (2000), 
and found a good agreement between the two in the redshift range where 
the $J-\Ks$ vs $R-\Ks$ method is valid ($1 \le z \le 2$).
We have also shown by Monte Carlo simulations 
that our method works successfully over $0<z<4$.

We have found that 143 ($58 \%$) of the EROs in our sample 
belong to the OE class. 
The redshift distribution of these OEs has a peak at $z \sim 1$ 
and a sharp cutoff at $z=0.8$. 
The surface density and the redshift distribution of OEs in our sample 
are broadly consistent with those given by previous authors.
Among the OEs, 24 \% are fit 
by a spectrum composed of bulge and disk components 
with the $B$-band bulge-to-total luminosity ratio of $\le 0.9$. 
It is also found that the OEs have a wide range of colors 
at any redshift.
These findings suggest not only that 
OEs cannot be reproduced by a single, 
passive evolution model with a fixed $\zf$ 
but also that a significant fraction of OEs at $z \ge 0.8$ 
have a non-negligible amount of star formation.

A comparison of the observed (cumulative) 
surface density of OEs with 
predictions from passive evolution models with $\zf=5, 3, 2.5$, 
and $2$ shows that the $\zf=3$ and 2.5 models reproduce 
fairly well the observed counts 
while the $\zf=5$ model and $\zf=2$ model overpredict and underpredict, 
respectively, counts at faint magnitudes.
In order to see more clearly 
in what range of redshift elliptical galaxies 
obey passive evolution, 
we have derived rest-frame $B$-band luminosity functions (LFs)
of OEs at $z=1-1.5$ and $1.5-2.5$ separately.
We have found that while the LF at $z=1-1.5$ agrees 
with the local LF of ellipticals if a dimming of 1.3 mag 
from $z=1.25$ to the present epoch is assumed, 
the amplitude of the LF at $z=1.5-2.5$ is lower than 
that of the local LF by a factor of $\sim 3$ 
over the whole range of magnitude observed, 
meaning that a strong number evolution has occurred 
at $1.5 \le z \le 2.5$. 
Therefore, the matches of the $\zf=3$ and 2.5 models 
to the observed number counts are superficial.
Previous authors have found a strong decrease in the number 
density of morphologically classified early-type galaxies 
at $z \gsim 1.5$.
Thus, the majority of ellipticals seen at present appear not to 
have established the two main features of ellipticals 
before $z \sim 1.5$, i.e., a red color and a smooth $1/4$-law profile. 

We have found that the angular correlation function of OEs 
shows a strong clustering, 
with $A_\omega = 6.1 \pm 0.9$ arcsec$^{0.8}$. 
We have estimated the spatial correlation length for OEs to be 
$r_0 = 11 \pm 1 h^{-1}$ Mpc, which is larger than that for 
the present-day early-type galaxies of similar luminosities.
This suggests that OEs are selectively located in regions 
which will become present-day clusters or groups of galaxies.

Finally, we have examined properties of the 104 DSs in our data.
We have found that the DSs have a wide range of redshifts, 
with a peak of the distribution at $z \sim 1.5$, 
and $E(B-V) \sim 0.4-1.2$ 
with a median of $\simeq 0.8$.
These redshift and $E(B-V)$ distributions are consistent 
with previous estimates.
The dust-corrected $SFR$ values are 
found to span $10^{1-3}$ $M_\odot$ yr$^{-1}$.
These high $SFR$s and a strong clustering seen in our DS sample 
may suggest that DSs are progenitors of part of 
the present-day E/S0s.


\acknowledgments
We would like to thank the Subaru Telescope staff
for their invaluable help in commissioning the Suprime-Cam
that makes these difficult observations possible.
The SIRIUS project was initiated and supported by Nagoya
University, National Astronomical Observatory of Japan, 
and University of Tokyo under a financial support of
Grant-in-Aid for Scientific Research on Priority Area (A)
No. 10147207 of the Ministry of Education, Culture, Sports,
Science, and Technology of Japan.
H. Furusawa, C. Nagashima, T. Nagayama, Y. Nakajima, 
F. Nakata, and M. Ouchi
acknowledge support from the Japan Society for the
Promotion of Science (JSPS) through JSPS Research Fellowships
for Young Scientists.


\appendix

\section{Method of Classification of EROs}

The method of our classification is a variant of the photometric 
redshift technique.
We use stellar population synthesis models by KA97 
to make model spectra.
KA97 include the chemical evolution of gas and stellar populations.
KA97 models have been successfully used to obtain photometric 
redshifts of the Hubble Deep Field North galaxies 
(Furusawa et al. 2000; 
see also Kodama et al. 1999 for an application 
to low-redshifts). 

In the present analysis, we use only a limited number of models
which are appropriate for EROs.
The model spectra for OEs contain 
not only spectra of passively evolving populations 
(i.e., bulge component) 
but also combined spectra of a bulge component and a disk 
component (see below) with various bulge-to-total luminosity 
ratios in the $B$ band, $B/T$, as shown in Table \ref{tab:SEDmodel}.
The reason for adding models with a disk component 
is to express a star formation which may have occurred 
in some OEs at high redshifts.
For bulges, we adopt $x=1.10$, $\tau_{\rm SF}=0.1$Gyr, 
$\tau_{\rm infall}=0.1$Gyr, and $t_{\rm GW}=0.2$Gyr, 
which are known to reproduce the average color of massive 
ellipticals in clusters of galaxies (Kodama et al. 1998).
For disks, we adopt $x=1.35$, $\tau_{\rm SF}=5$Gyr, 
$\tau_{\rm infall}=5$Gyr, and $t_{\rm GW}=20$Gyr 
(this means that the galactic wind does not blow by the present-day), 
which are close to the values estimated for the disk of our Galaxy.
We make composite spectra 
by combining a disk component and a bulge component 
with the same age to give 18 combinations of different $B/T$.
The range of ages for OE models is set to 
$1 \le {\rm age}\hspace{3pt}({\rm Gyr}) \le 15$.
We adopt very small intervals of $B/T$ 
at $B/T \ge 0.9$ (see Table \ref{tab:SEDmodel}) 
in order to completely cover the region in color-color spaces 
occupied by the observed EROs with a sufficiently small grid. 
Note that adding a small amount of disk component makes 
colors of the composite spectrum 
in short wavelengths, such as $B-R$, drastically blue.

The model spectra for DSs are generated by adding dust 
extinction to young disk spectra whose parameters are 
the same as for disks used for OEs (Table \ref{tab:SEDmodel}).
However, the maximum age is set to $\le 1.585$ Gyr. 
The range of dust extinction is set to $0 \le E(B-V) \le 1.5$ 
with an interval of $E(B-V)$ of 0.05.
We adopt the extinction curve given in Calzetti (1997).

We redshift these model spectra from $z=0$ to $4$ 
with an interval of 0.05, to obtain a full set of model spectra 
for classification of EROs.
For each spectrum, we take account of the effect of absorption 
due to the intergalactic HI gas, following the prescription 
by Madau (1995); this effect is important 
only for galaxies at $z \gsim 4$.
Finally, we convolve each redshifted model spectrum by the system
response functions, and obtain a set of 
{\lq}model fluxes{\rq} from $B$ to $\Ks$.

For each ERO, we compare the convolved model flux $T_{i}$ 
with the observed flux $F_{i}$ in
the $i$-th bandpass and calculate $\chi^2$ as:
\begin{eqnarray}
 \chi^2 &=& \sum_{i}^{N_{band}}\frac{(F_i - \alpha\,T_i)^2}
                                     {\sigma_i^2},\label{eq:chi2}
\end{eqnarray}
\noindent
where $N_{band}$ is the number of bands used for the observation 
and $\sigma_i$ is an observational error in the $i$-th band.
For a given set of $T_i$, the value of $\alpha$ which minimizes
$\chi^2$ is calculated as:
\begin{eqnarray}
\Deriv{\chi^2}{\alpha} &=& \Deriv{\sum_{i}^{N_{band}} 
  \frac{(F_i - \alpha\,T_i)^2}{\sigma_i^2}}{\alpha} = 0,
  \label{eq:calc_alpha}\nonumber\\
  \alpha &=& \frac{\sum_{i}^{N_{band}} 
  \frac{F_i T_i}{\sigma_i^2}}{\sum_{i}^{N_{b
and}} \frac{T_i^2}{\sigma_i^2}}.\label{eq:alpha}
\end{eqnarray}
Then, we find the best-fit spectrum which gives 
`the minimum reduced $\chi^2$'
out of all the model spectra, 
and the class and the redshift of the best-fit spectrum are
adopted as those of the ERO.
At the same time, the best-fit $E(B-V)$ value and the age are derived.

\begin{figure*}
\epsscale{1.0}
\plotone{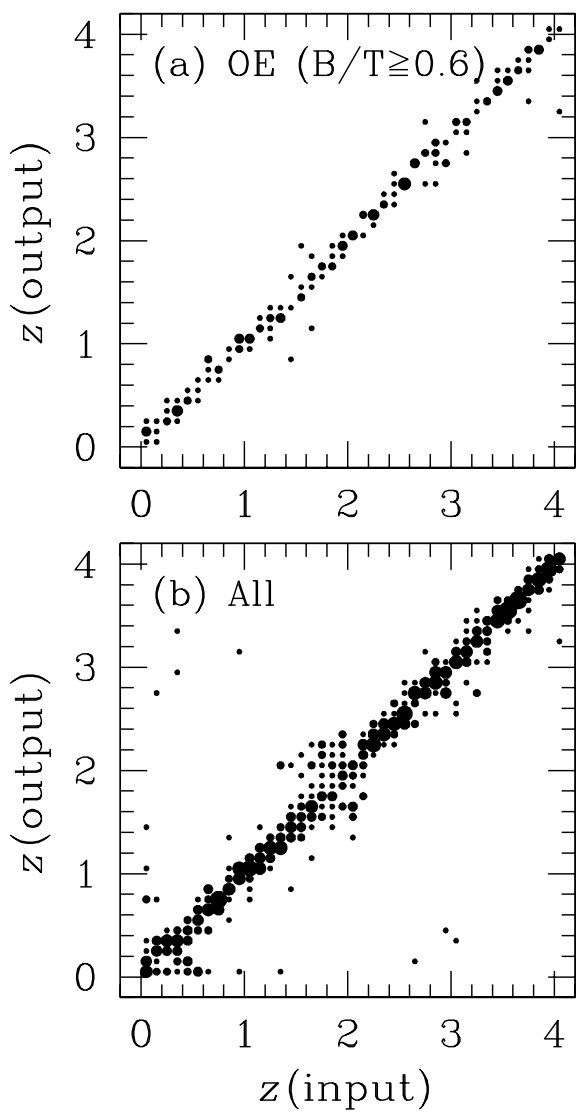}
\caption{Simulations of photometric redshift estimation 
         using model spectra.
         Abscissa is the redshift of an input spectrum,
         and ordinate shows the redshift determined from 
         the spectrum fitting to the input spectrum.
         The area of the filled circles is proportional to 
         the number of objects in the grid.
         Panel (a): 
         Randomly selected 153 OEs with $B/T$(input) $\ge$ 0.6.
         Panel (b): 
         Randomly selected 530 spectra including both OEs and DSs.
    \label{fig:photz_simul}}
\end{figure*}

To check the reliability of this method, 
we carry out Monte Carlo simulations.
We randomly select 530 galaxies 
from the set of model spectra over $z=0-4$, add to them 
a $10\%$ noise (in each bandpass), which is close to 
the typical photometric errors for our EROs,
and fit the template model spectra to them 
in the same manner as for the observed EROs.
As we mention in the main text, all OEs in our data have $B/T \ge 0.6$.
Thus, in order to examine uncertainties in redshift estimates 
for OEs, we compare in Figure \ref{fig:photz_simul} (a) 
the redshifts of input spectra with the estimated redshifts 
for OEs whose input $B/T$ is $\ge 0.6$.
We find that objects with $|z({\rm output})-z({\rm input})| > 0.5$ 
are very few, and that the rms scatter around the equality is 0.14 
even when no object is clipped.
The success rate of classification is found to be high; 
the fraction of input OEs with $B/T\ge 0.6$ which are
correctly classified as OEs with $B/T\ge 0.6$ by the fit is $92\%$.
For the reader's reference, Figure \ref{fig:photz_simul} (b) 
compares the redshifts of input spectra 
with the estimated redshifts for the all (530) galaxies.
A tight correlation is found between the input and output 
redshifts, although there are a small fraction of galaxies 
with a catastrophic error in redshift estimate; 
the rms scatter around the equality is 0.14
if galaxies with $|z({\rm output})-z({\rm input})| > 0.5$
are removed.

About 25 \% of the EROs in our sample do not have $J$ data.
To examine effects of not using $J$ flux on the fit, 
we make similar Monte Carlo simulations without using $J$ flux. 
We find that the performances of classification and redshift 
estimation do not change significantly.




\clearpage


\clearpage

\clearpage


\begin{deluxetable}{crcc}
\footnotesize
\tablecaption{Summary of the observational data
\label{tab:Obsdata}
}
\tablewidth{0pt}
\tablehead{\colhead{optical data} & \colhead{ } & 
\colhead{ } & \colhead{ }
}
\startdata
 & \multicolumn{3}{l}{coverage: 618 arcmin$^2$}\\
 & \multicolumn{3}{l}{PSF FWHM: $0''.98$}\\
\tableline
Band & Exp.[min] & $m_{\rm lim}^{(a)}$ & Obs.Date \\
\tableline
$B$ & 177 & 27.1 & Nov 2000 \\
$V$ & 108 & 25.9 & Nov 2000 \\
$R$ & 138 & 26.3 & Nov 2000, Nov 2001 \\
$i'$ & 45 & 25.7 & Nov 2000 \\
$z'$ & 40 & 25.0 & Oct 2001 \\
\tableline
\tableline
\multicolumn{4}{l}{infrared data}\\
\tableline
 & \multicolumn{3}{l}{coverage: 114 arcmin$^2$ ($H, \Ks$), 
                                77 arcmin$^2$ ($J$)}\\
 & \multicolumn{3}{l}{PSF FWHM: $0''.98$}\\
\tableline
Band & Exp.[min] & $m_{\rm lim}^{(a)}$ & Obs.Date \\
\tableline
$J$   & 120 & 22.8 & Aug, Sep 2001 \\
$H$   & 120 & 22.5 & Aug, Sep 2001 \\
$\Ks$ & 120 & 22.1 & Aug, Sep 2001 \\
\tableline
\enddata
\tablenotetext{(a)}{$5\sigma$ on a $2''$ diameter aperture}
\end{deluxetable}

\clearpage

\begin{deluxetable}{lr}
\footnotesize
\tablecaption{$B/T$ distribution of OEs
\label{tab:BTdist}
}
\tablewidth{0pt}
\tablehead{ 
 \colhead{$B/T$} & \colhead{number} 
}
\startdata
0.6   &  1 \\
0.7   &  3 \\
0.8   &  9 \\
0.9   & 22 \\
0.95  & 38 \\
0.98  & 22 \\
0.99  & 13 \\
0.995 &  6 \\
0.996 &  2 \\
0.997 &  2 \\
0.998 &  0 \\
0.999 &  1 \\
1.0   & 24 \\
\tableline
\enddata
\end{deluxetable}

\clearpage

\begin{deluxetable}{lccccl}
\footnotesize
\tablecaption{Parameters of model spectra adopted in this work
\label{tab:SEDmodel}
}
\tablewidth{0pt}
\tablehead{
 \colhead{SED type} & \colhead{$x$\tablenotemark{(a)}} & 
 \colhead{$\tau_{\rm SF}$ [Gyr]} & \colhead{$\tau_{\rm infall}$ [Gyr]} & 
 \colhead{$t_{\rm GW}$ [Gyr]\tablenotemark{(b)}} & \colhead{Age[Gyr]}
}
\startdata
OE        & {}   & {}  & {}  & {}    & 1.0, 2.0, 3.0, 4.0, 5.0,\\
(pure bulge)& 1.10 & 0.1 & 0.1 & 0.2 & 6.0, 7.0, 8.0, 9.0, 10.0,\\
{}        & {}   & {}  & {}  & {}    & 11.0, 12.0, 13.0, 14.0,15.0\\
\tableline
OE                           & {}   & {}  & {}  & {}    & 
    1.0, 2.0, 3.0, 4.0, 5.0,\\
(bulge+disk)\tablenotemark{(c)}  & {$\cdots$}  & {$\cdots$} & 
   {$\cdots$} & {$\cdots$}   & 6.0, 7.0, 8.0, 9.0, 10.0,\\
{}& {} & {} & {}  & {}  & 11.0, 12.0, 13.0, 14.0, 15.0\\
\tableline
{}   & {}   & {}  & {}  & {}   &  0.010, 0.013, 0.016, 0.020, 0.025,\\
{}   & {}   & {}  & {}  & {}   &  0.032, 0.040, 0.050, 0.063, 0.790,\\
DS   & 1.35 & 5.0 & 5.0 & 20.0 &  0.100, 0.126, 0.158, 0.200, 0.251,\\
{}   & {}   & {}  & {}  & {}   &  0.316, 0.398, 0.501, 0.631, 0.794,\\
{}   & {}   & {}  & {}  & {}   &  1.0, 1.259, 1.585 \\
\tableline
\enddata
\tablenotetext{(a)}{Power-law index of initial mass function.}
\tablenotetext{(b)}{$t_{\rm GW}=20$ means that the galactic wind 
 does not blow until the present epoch.}
\tablenotetext{(c)}{$B/T=0.1$, 0.2, 0.3, 0.4, 0.5, 0.6, 0.7, 0.8, 
 0.9, 0.95, 0.98, 0.99, 0.995, 0.996, 0.997, 0.998, 0.999, 1, 
 where $B$ and $T$ are the rest-frame 
 bulge and the total luminosity in the $B$ band. 
The values of $x$, $\tau_{\rm SF}$, $\tau_{\rm infall}$, 
and $t_{\rm GW}$ for disks are the same as for DS types.}
\end{deluxetable}

\clearpage

\end{document}